\documentclass[preprint,11pt, 3p]{elsarticle}

\usepackage[bookmarks=true, pdfpagelabels=true, colorlinks=true, linkcolor=black, urlcolor=black, citecolor=blue, anchorcolor=blue]{hyperref}
\usepackage[T1]{fontenc} 
\usepackage[dvipsnames]{xcolor}
\usepackage{slashed}
\usepackage{enumitem}
\usepackage{amsmath}
\usepackage{amssymb}

\newcommand{\mgFull}{{\sc MadGraph5\_aMC@NLO}~}

\journal{Physics of the Dark Universe}

\begin{document}

\begin{frontmatter}

\title{Circular polarisation of gamma rays as a probe of dark matter interactions with cosmic ray electrons}

\author[]{Marina Cermeño}
\ead{marina.cermeno@uclouvain.be}
\author[]{Céline Degrande}
\ead{celine.degrande@uclouvain.be}
\author[]{Luca Mantani}
\ead{luca.mantani@uclouvain.be}

\address{Centre for Cosmology, Particle Physics and Phenomenology (CP3), \\
Universit\'e Catholique de Louvain, \\
B-1348 Louvain-la-Neuve, Belgium}

\null\hfill\begin{tabular}[t]{l@{}}
  CP3-21-08 \\
  MCNET-21-04
\end{tabular}

\begin{abstract}
Conventional indirect dark matter (DM) searches look for an excess in the electromagnetic emission from the sky that cannot be attributed to known astrophysical sources. Here, we argue that the photon polarisation is an important feature to understand new physics interactions and can be exploited to improve our sensitivity to DM. In particular, circular polarisation can be generated from Beyond the Standard Model interactions if they violate parity and there is an asymmetry in the number of particles which participate in the interaction. In this work, we consider a simplified model for fermionic (Majorana) DM and study the circularly polarised gamma rays below $10$ GeV from the scattering of cosmic ray electrons on DM. We calculate the differential flux of positive and negative polarised photons from the Galactic Center and show that the degree of circular polarization can reach up to $90\%$. Finally, once collider and DM constraints have been taken into account, we estimate the required sensitivity from future experiments to detect this signal finding that, although a distinctive peak will be present in the photon flux spectrum, a near future observation is unlikely. However, different sources or models not considered in this work could provide higher intensity fluxes, leading to a possible detection by e-ASTROGAM. In the event of a discovery, we argue that 
the polarisation fraction is a valuable characterisation feature of the new sector.
\end{abstract}


\end{frontmatter}


\section{Introduction}
\label{sec:intro}

The evidence of the existence of Dark Matter (DM), a form of unknown matter that constitutes approximately 85\% of the matter content in  the  presently  accepted cosmological model for our Universe~\cite{Planck:2020Id0}, is supported by many astrophysical and cosmological observations~\cite{zwicky1, zwicky2, vera, bulletcluster1, bulletcluster2, bc3, cmb, cmb1, cmb2}. Despite the impressive experimental efforts of the last few decades, its identity is still unknown (see ref.~\cite{bertone:2016} for a review).

One of the most popular hypothesis is that DM is constituted of Beyond Standard Model (BSM) particles. Under this assumption, different experimental strategies have been implemented and we are arguably entering in an era where the precision of experiments is getting high enough to reach conclusive statements. One of these search strategies concerns indirect detection of DM through their  annihilation  or  decay  into  SM  products in high density regions, such as the Galactic Center (GC)~\cite{vanEldik:2015qla} or nearby Dwarf Spheroidal Galaxies~\cite{Strigari:2018utn}.
The main obstacle for this kind of searches is the difficulty to  distinguish possible signals from astrophysical backgrounds. 

In this context, the observability of peaks in the astrophysical spectrum, such as monochromatic lines or similar features related to internal bremsstrahlung, provide smoking gun signatures. 
For example, the presence of peaks in the gamma-ray spectrum due to the annihilation of fermionic Majorana DM has been thoroughly studied~\cite{Bringmann:2012vr, Garny:2013, Kopp:2014, Okada:2014zja, Garny:2015wea, Kumar:2016cum, Bartels:2017dpb}. In this scenario, the 2 to 2 annihilation of DM via a charged mediator is p-wave suppressed and internal bremsstrahlung is more likely to happen. In addition, the signal due to this process for a mediator close in mass to the DM particle provides a sharp peak in the spectrum which can be found at the DM mass energy scale. 
While DM coupling to quarks is already tightly constrained~\cite{Garny:2015wea}, the leptophilic case still offers a wide window to explain the DM nature.

On the other hand, an important signature from this class of models, which has been analysed in ref.~\cite{Gorchtein:2010xa, Profumo:2011jt, Huang:2011dg, Gomez:2013qra}, is the peak in the spectrum expected at the DM-mediator mass splitting energy due to the scattering of DM by cosmic ray (CR) electrons.  
The number of photons at this energy can, depending on the parameters of the model (mass splitting, coupling, etc), be of the same order than the ones corresponding to the annihilation peak. We will elaborate on this later. Nevertheless, what can certainly differentiate the two signals is the fact that only the flux of photons coming from the DM-CR electron scattering can be circularly polarised.

Recent works~\cite{Kumar:2016cum,Elagin:2017cgu,Gorbunov:2016zxf,Boehm:2017nrl, Huang:2019ikw, Balaji:2019fxd, Balaji:2020oig} suggest that DM and neutrinos can generate circular polarised signals in X-rays or gamma-rays through decays and interactions  with SM particles, pointing out an unexplored way
to look for new physics. However, the circular polarisation of X-ray and gamma-ray photons has been proposed much earlier as a signature of SM parity (P) violating neutrino interactions in the outer layer of a collapsing stars~\cite{khlopov}. A net circular polarisation signal is generated when there is an excess of
one polarisation state over the other. As circular polarisation states flip under parity, P must be violated in at least one of the dominant photon emission processes. Moreover, there must be either an asymmetry in the number density of one of the particles in the initial
state or CP must be violated by the interactions at play. Therefore, even if a process violates P, circular polarisation is only possible when the initial state is not a CP-eigenstate. In particular, DM annihilation cannot produce polarised photons unless CP violating interactions are present. This is explained in more detail in Section~\ref{sec:circular_polarisation}. In this regard, in~\cite{Boehm:2017nrl}, the fraction of circular polarisation coming from the interaction between electrons and DM in a region with an excess of electrons over positrons has been analysed by calculating the cross section for an electron scattering off a neutralino radiating a positive or a negative circularly polarised photon, finding that the difference between both cross sections can be significant. This suggests that a net circular polarisation signal can be generated by DM interactions with ambient CRs. Motivated by this, here we perform for the first time the full computation of the flux of circularly polarised photons coming from these kind of interactions in the GC, taking into account the energy distribution of CR electrons and the DM density. After finding that the circular polarisation fraction can reach values higher than $90\%$, we argue that this feature and their energy dependence could be used to reveal these BSM interactions as well as to study their nature.

So far, measurements of circular polarisation signals have been achieved in an astrophysical context for the radio energy band~\cite{Han:1998zk,Han_2009} as a probe of synchrotron emission~\cite{deBurca:2015kea}. For the case of gamma-rays, circular polarization can be measured via Compton scattering, exploiting the correlation of the outgoing electron spin with the initial photon helicity. Nevertheless, although future existing missions, like e-ASTROGRAM~\cite{DeAngelis:2017gra} or AMEGO~\cite{McEnery:2019tcm}, will be sensitive to linear polarization in the $1$ MeV to $1$ GeV energy band (see~\cite{Ilie:2019yvs} for a review on this), they are not designed to measure circular polarisation in this energy range. 
The possibility of observation is now an open question, but a great upside could encourage the experimental community to explore the feasibility of such detectors in the near future.

In the following we will consider a simplified model where the DM is a Majorana fermion which
couples to SM leptons via a charged scalar mediator. Taking into account their interactions with CR electrons in the GC, we will study the flux of circularly polarised photons radiated in the final state for a scenario where the mediator and the DM are close in mass.

The focus of this analysis is devoted in arguing that this approach opens the way to exploiting the photon polarisation as a feature to understand new physics interactions.


The paper is structured as follows. In Section~\ref{sec:circular_polarisation} we discuss how circular polarisation can be generated given a particular interaction. In Section~\ref{sec:model} we introduce the model we consider and we show the existing constraints given by the measured relic density and the bounds of direct, indirect and collider experiments. In Section~\ref{sec:flux_results} we describe the calculation we have carried out for the flux of circularly polarised photons and we present our results. In Section~\ref{sec:detection} we analyse the detection prospects and, finally, in Section~\ref{sec:conclusions} we summarize our conclusions.

\section{Circular polarisation}
\label{sec:circular_polarisation}

In order to have a source of net circularly polarised photons, the underlying physics has to violate both P and CP. In this section we discuss this aspect and present the convention used for the photon polarisation vectors.

The need of P violation arises from the fact that the photon flips helicity under a P transformation. If one of the particles in the initial or final state prefers one particular helicity, i.e. the coupling is stronger to one polarisation state than to the other, that means that P is violated. This happens, for example, when the initial or final state is purely left or right-handed. Nevertheless, this is not the only condition required. If P is violated but CP is not, then the CP-conjugate process will provide the same rate for producing the opposite helicity, unless there is an asymmetry in the number density of one of the particles in the initial state, i.e. their number density is not the same as the number density of its antiparticle. An example of the latter is the P violating interaction that we consider in this work for the DM electron scattering $\tilde{\chi} e^- \rightarrow \tilde{\chi} e^- \gamma$. In this case, although the interaction conserves CP, as the density of electrons is higher than the one of positrons in the region where the interaction takes place, a circular polarisation signal can be generated. On the contrary, we will not expect this signature from the annihilation, $\tilde{\chi} \tilde{\chi} \rightarrow e^- e^+ \gamma$, since the initial state is a CP-eigenstate. 

In this work, we will use the same convention for the photon polarisation vectors that is used in~\cite{Boehm:2017nrl}. We consider a photon with momentum $k^\mu=(k_0, k_x, k_y, k_z)$, whose two possible transverse polarisation vectors are
\begin{equation}
\epsilon_1^\mu(k)=\frac{1}{k_0 k_T}(0, k_x k_z, k_y k_z, -k_T^2)    
\end{equation}
and
\begin{equation}
\epsilon_2^\mu(k)=\frac{1}{k_T}(0, -k_y, k_x, 0), 
\end{equation}
where $k_T=\sqrt{k_x^2+k_y^2}$. If the photon is travelling along the z-direction, i.e. $k^\mu=(k_0, 0, 0, k_z)$ and $k_T=0$, then 
\begin{equation}
\epsilon_1^\mu(k)=(0,1, 0, 0),    
\end{equation}
\begin{equation}
\epsilon_2^\mu(k)=(0, 0, \frac{k_z}{|k_z|}, 0).
\end{equation}
The positive and negative photon circular polarisation vectors are then defined as
\begin{equation}
\epsilon_{\pm}^\mu(k)= \frac{1}{2}(\mp\epsilon_1^\mu(k)-i \epsilon_2^\mu(k) ).   
\end{equation}

These definitions allow us to define the squared helicity amplitudes $\mathcal{A}_-= \sum_{spins} |\epsilon^{\mu}_{-}\mathcal{M}_\mu|^2$ and $\mathcal{A}_+= \sum_{spins} |\epsilon^{\mu}_{+}\mathcal{M}_\mu|^2$, where $\mathcal{M}_\mu$ is the leftover of the Feynman amplitude after factorising the polarisation vector, such that their sum is equivalent to the total averaged amplitude. For a parity violating interaction $\mathcal{A}_- \neq \mathcal{A}_+$ and a net circularly polarised spectrum is expected.

\section{Dark Matter observables and existing  constraints}
\label{sec:model}

\begin{figure}[t!]
  \centering
  \includegraphics[scale=0.23]{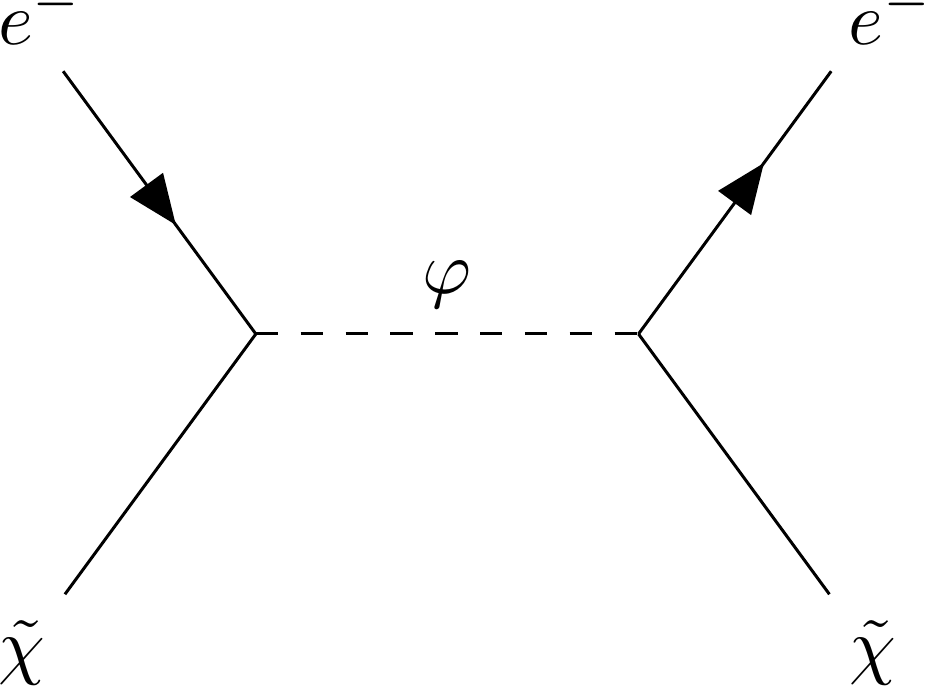}\\
  \vspace{0.5cm}
  \includegraphics[scale=0.23]{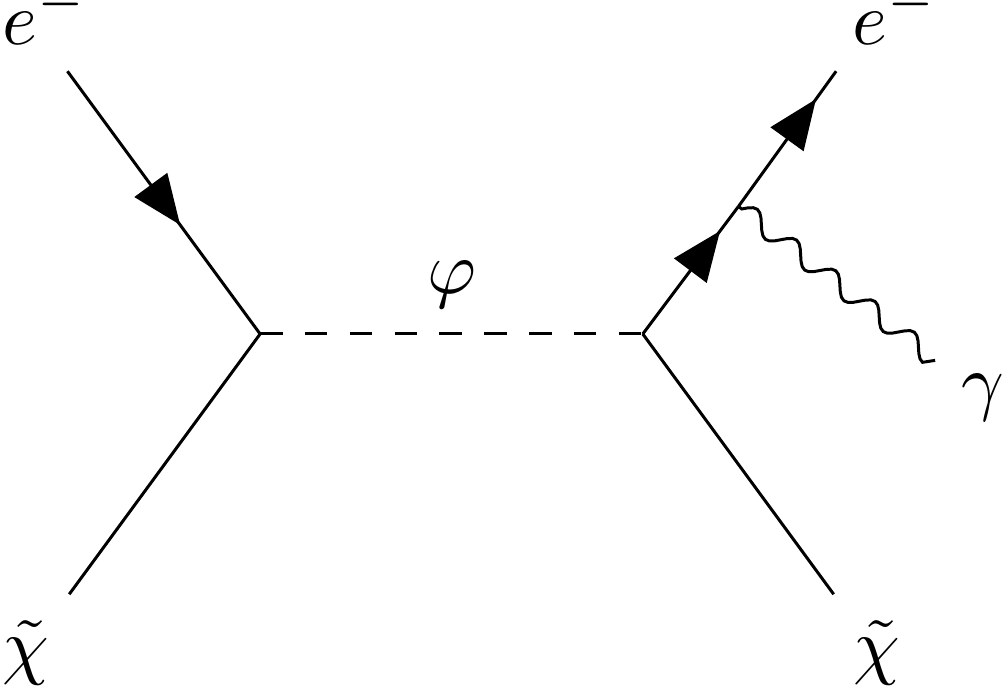} \qquad
  \includegraphics[scale=0.23]{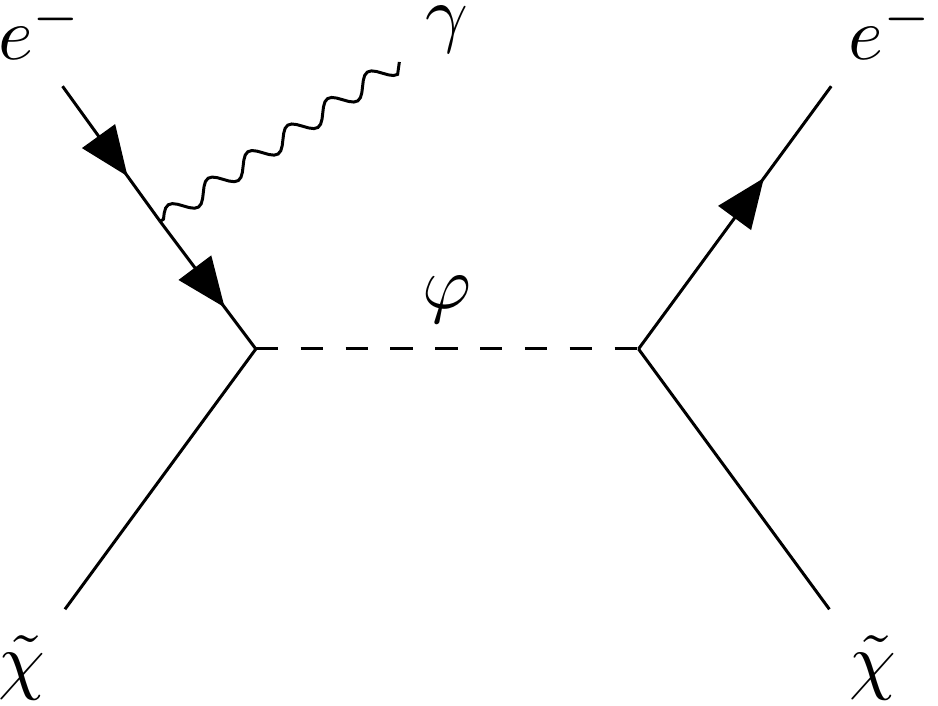} \qquad
  \includegraphics[scale=0.23]{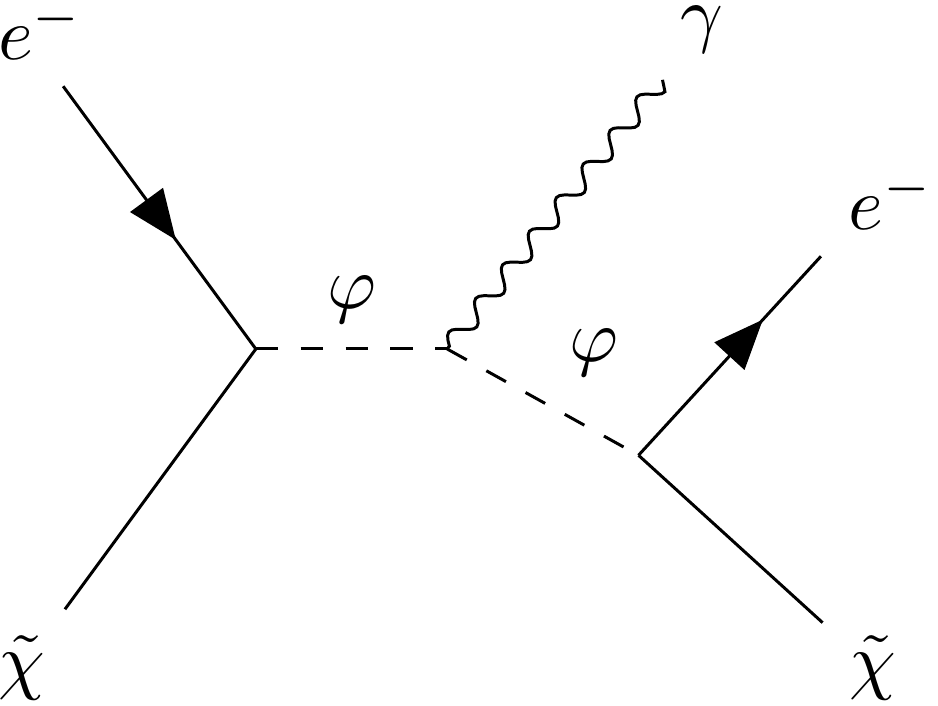}
  \caption{
\label{fig:diagrams}
Diagrams for relevant production modes to this study of electrons scattering with a Majorana DM particle through a right-handed scalar mediator. In the upper part we show the 2 to 2 scattering process. In the lower part the resonantly enhanced 2 to 3 radiative process. }
\end{figure}

As long as DM interacts with the SM through a P violating interaction, there is the possibility that a distinctive polarised signal is produced by the scattering with CR. 
As an example, in order to provide quantitative results and estimates, in this work we consider a specific leptophilic simplified DM model. In particular, following~\cite{Profumo:2011jt, Kopp:2014, Garny:2015wea}, we study a scenario in which the SM lagrangian is extended with two additional degrees of freedom, a neutral fermion, $\tilde{\chi}$, and a charged scalar mediator, $\varphi$, coupled to leptons. Specifically, we consider the t-channel model in which the DM particle is coupled to right-handed electrons, $e_R$, with a coupling constant $a_R$. The lagrangian for the dark sector reads

\begin{equation}
    \mathcal{L}_{DM} = i \bar{\psi}_{\tilde{\chi}}(\slashed{D} - m_{\tilde{\chi}})\psi_{\tilde{\chi}} + D_\mu \varphi^\dagger D^\mu \varphi - m_\varphi \varphi^\dagger \varphi + ( a_R \, \bar{e}_R \, \psi_{\tilde{\chi}} \, \varphi + h.c.) \, .  
\end{equation}

This simplified scenario can be understood in a supersymmetry (SUSY) context where $\tilde{\chi}$ is identified with the lightest neutralino (a Majorana particle), and $\varphi$ with the right-handed selectron of the Minimal Supersymmetric Standard Model (MSSM). The scalar mediator is carrying the same quantum numbers of the right-handed electron, since the DM is a singlet of the SM gauge groups. The model parameter space is three dimensional, i.e. is characterised uniquely by the set of parameters $\{m_{\tilde{\chi}}, m_\varphi, a_R \}$. Additionally, in order to ensure DM stability, the mass of the mediator has to be bigger than the mass of the DM candidate ($m_\varphi > m_{\tilde{\chi}}$). Note that we do not consider couplings to left-handed electrons since they are more strongly constrained, see for example~\cite{Liu:2013gba}. Moreover, a pure right-handed (or left-handed) coupling will provide a maximum circular polarisation fraction.

The choice of considering a t-channel model stems from the fact that in this class of models, DM-SM scatterings can be kinematically enhanced by a resonant contribution. In particular, if the mass splitting between DM and the mediator, $\Delta M = m_\varphi - m_{\tilde{\chi}}$, is small, one can think of exploiting the mediator resonance in scattering of CR electrons (see Fig.~\ref{fig:diagrams}). This kind of scattering can supplement the indirect detection efforts and open a window to probe the mass degenerate region of the parameter space, which is hard to probe with other methods, as we will see in the following.

In order to compute DM observables we have implemented our model in FeynRules~\cite{Alloul:2013bka}, which allowed us to export the model in formats suitable for tools like FeynArts~\cite{Hahn:2000kx}, FeynCalc~\cite{feyncalc1, feyncalc2, feyncalc3},  \mgFull~\cite{Alwall:2014hca}, MadDM~\cite{Ambrogi:2018jqj, Arina:2020kko} and MicrOMEGAs~\cite{Belanger:2018ccd}.

In this section we want to briefly report on the current limits posed by experiments on the parameter space of the model under study. In particular, we consider the case of a Majorana DM, being well-motivated by numerous BSM scenarios. Being interested in the region of the mass spectrum that allows us to exploit the resonant production efficiently, we focus only on small mass splitting $\Delta M \sim [10^{-3}, 10]$ GeV.

\subsection{Relic density}
\label{subsec:relic}

If DM is produced thermally, we can predict the expected relic density for a specific point of the parameter space. In particular, for each point in the 2D mass plane ($m_{\tilde{\chi}}$, $\Delta M$), we can find the value of the coupling, $a_R$, that yields the Planck measurement~\cite{Planck:2020Id0} for the relic abundance ($\Omega_{\chi}h^2= 0.120 \pm 0.001$). We computed the relic density taking into account both annihilation and coannihilation, which is particularly relevant when $r = m_\varphi / m_{\tilde{\chi}} \leq 1.2$, see~\cite{Garny:2015wea, Baker:2018uox, Junius:2019dci}. 

\begin{figure}[t!]
  \centering
  \includegraphics[width=.48\columnwidth]{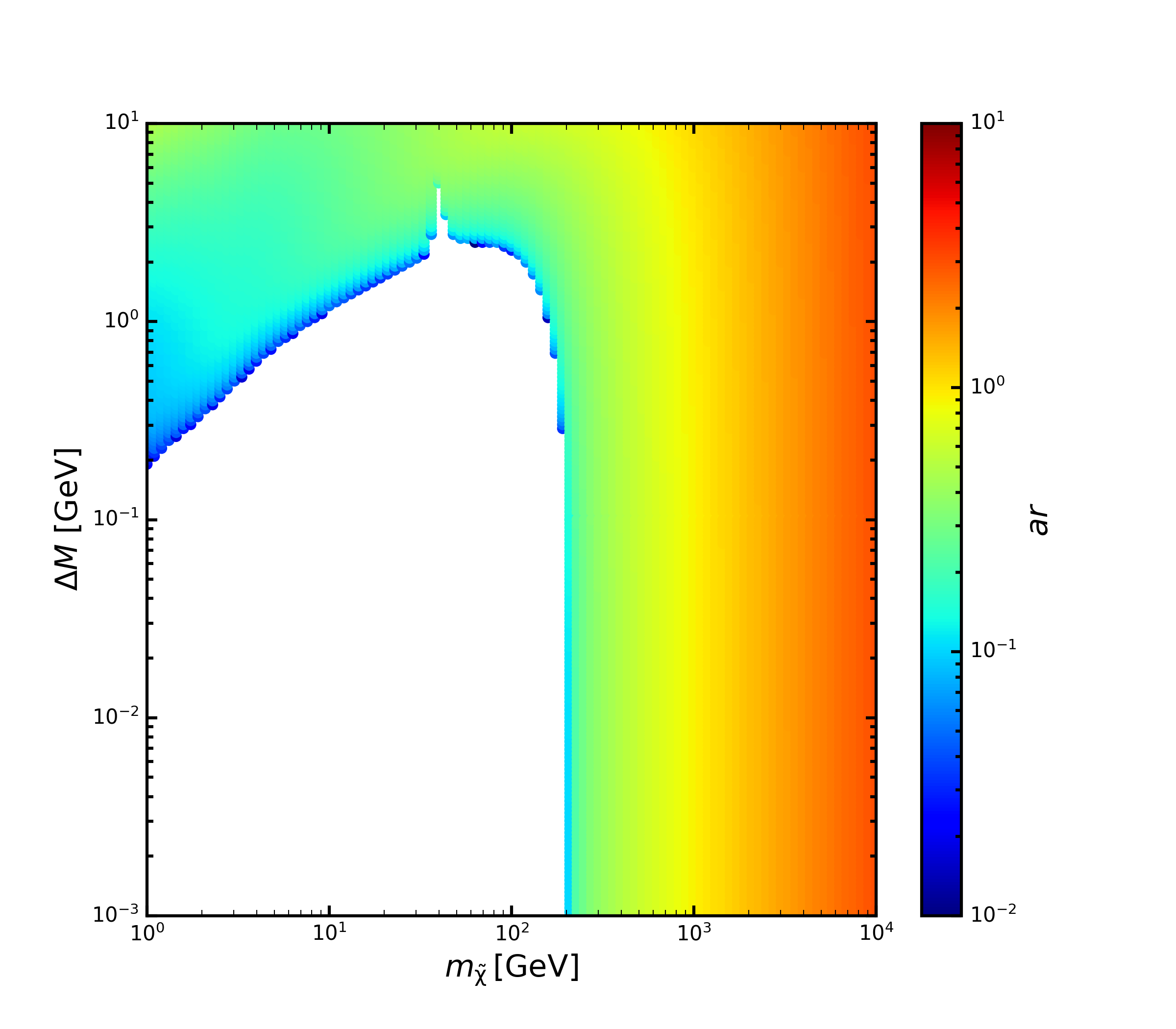}
  \includegraphics[width=.48\columnwidth]{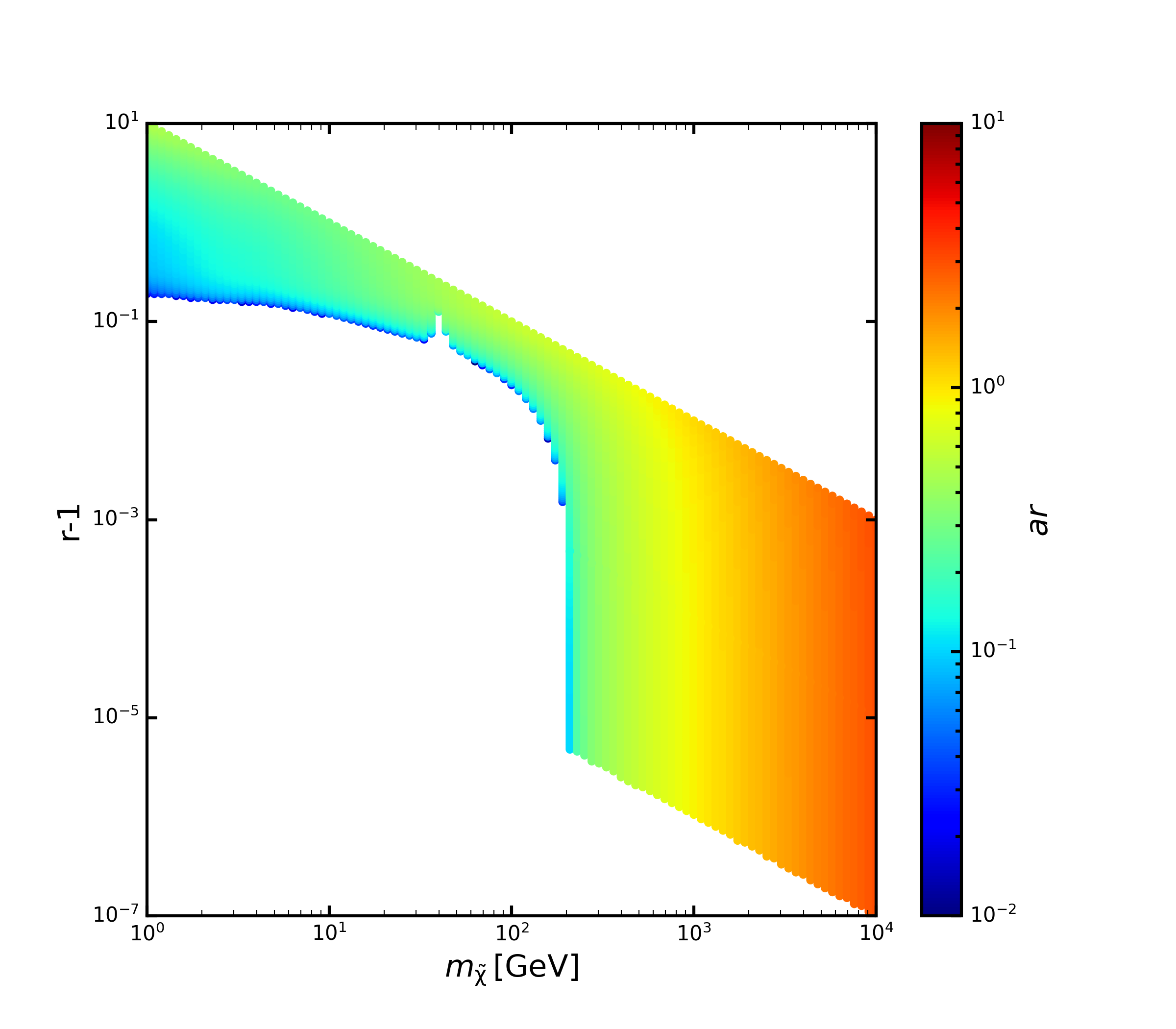}
  \caption{
    \label{fig:relic}
    Parameter space points that yield the correct relic density measured by Planck $\Omega_{\chi}h^2= 0.120 \pm 0.001$. On the left, the points in the $\Delta M - m_{\tilde{\chi}}$ plane while on the right the same points with $r-1 = m_\varphi/m_{\tilde{\chi}} - 1$ on the y-axis to better appreciate the compressed region.}
\end{figure}

In Fig.~\ref{fig:relic} we have plotted the parameter space points which give a relic abundance compatible with the currently observed value. We show the solutions both in the $\Delta M - m_{\tilde{\chi}}$ and the $(r-1) - m_{\tilde{\chi}}$ planes. 

A notable feature that can be seen from the plots is the fact that in the area at small $\Delta M$, $r-1 \lesssim 10^{-1}$, and $m_{\tilde{\chi}} \lesssim 100$ GeV the relic density is completely dominated by the coannihilation of the mediator, which is independent of the coupling $a_R$ since it is a pure electroweak (EW)  process, see for example~\cite{Garny:2015wea, Junius:2019dci}. This translates into the fact that solutions in this region are not available, since coannihilation is so efficient that the DM relic density observed by Planck cannot be produced via freeze out. On the other hand, as we look at higher DM masses, coannihilation cross sections start to become smaller and the space of solutions opens up again beyond $m_{\tilde{\chi}} \gtrsim 200 \, \mathrm{GeV}$. In this region, given the restricted range $\Delta M \sim [10^{-3}, 10] \, \mathrm{GeV}$ chosen, we are always in coannihilation regime but in order for the relic density to be the one observed by Planck, DM annihilation plays a role as well. In particular, the higher the DM mass, the bigger the coupling is needed to compensate the loss in cross section from coannihilation. For a fixed DM mass, we find that the relic abundance in this parameter space region does not depend on the value of $\Delta M$, since we are always in a highly degenerate case, see the plot in $(r-1)-m_{\tilde{\chi}}$ plane in Fig.~\ref{fig:relic}. Notice that the absence of points in the upper part of the plot is only due to the fact that we restricted the scan to maximum $\Delta M = 10$ GeV, but they are perfectly viable parameter space points mostly dominated by DM annihilation. On the other hand, the effect of the Z pole can be observed for $m_{\tilde{\chi}} \sim 45$ GeV due to the opening of the resonant contribution $\varphi^\dagger \varphi \to Z \to ll/qq $ for small mass splitting. Because of this resonant enhancement to the coannihilation cross section, there is an absence of solutions in that parameter space region and a DM under abundance.

\subsection{Indirect detection}
\label{subsec:indirect}
\begin{figure}[!t]
  \centering
  \includegraphics[width=.6\columnwidth]{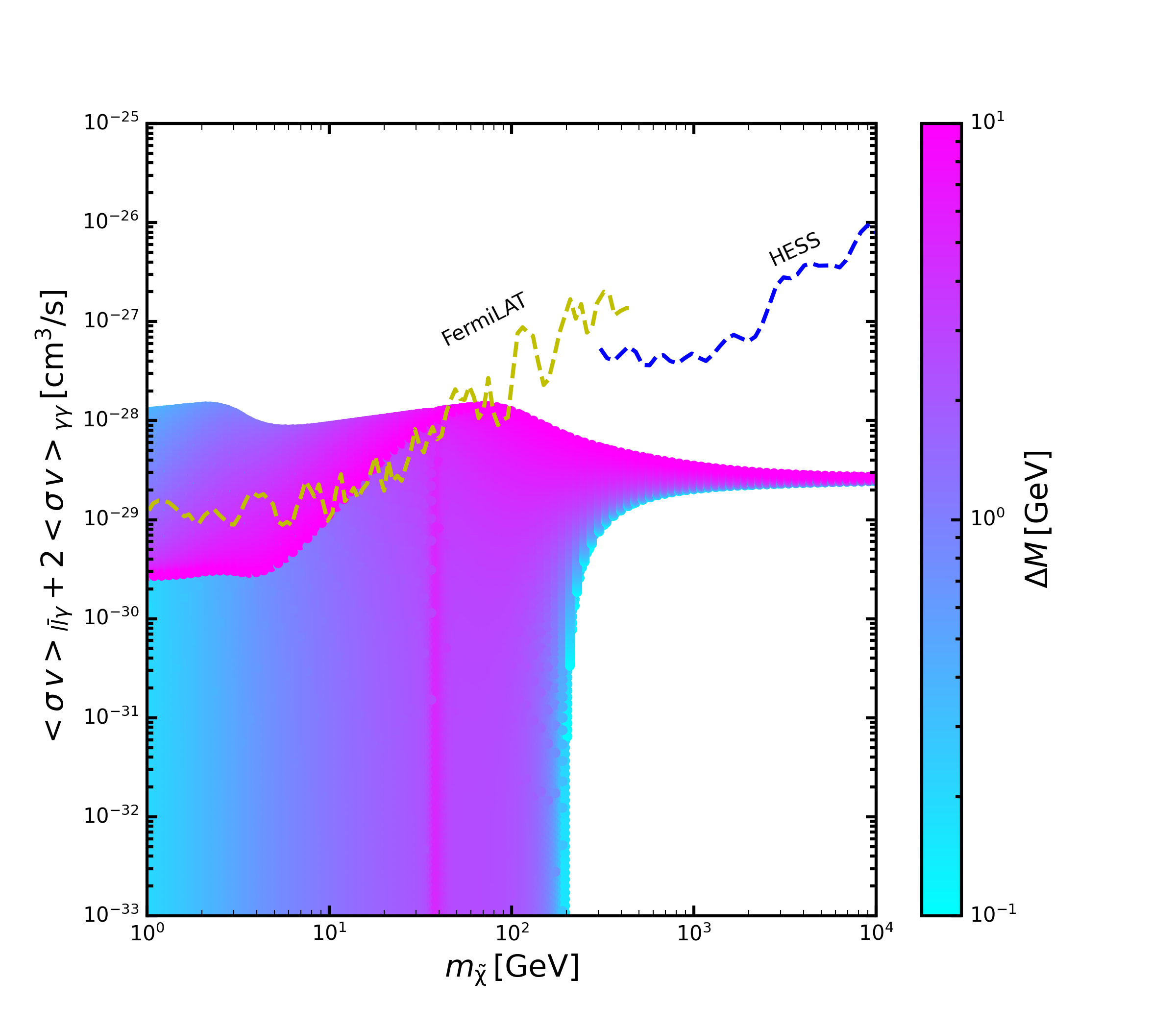}
  \caption{
    \label{fig:indirect}
    Averaged annihilation cross section as a function of the DM mass for different values of the mass splitting which provide the measured relic abundance. The yellow and blue lines are the upper limits from Fermi-LAT~\cite{fermi:2015lka} and HESS~\cite{hess:2018qtu} for this quantity by considering an Einasto DM density profile~\cite{einasto:1989,Navarro:2008kc, Pieri:2009je}.}
\end{figure}

Under the hypothesis that no mechanism produced an asymmetry between DM and anti-DM abundance or DM is self-conjugate, which is the case of our Majorana DM candidate, pair annihilation in high density regions generates a possibly detectable signal in the gamma-rays that would reach our telescopes on Earth. In particular, the main contribution is usually given by $\tilde{\chi} \tilde{\chi} \to e^+ e^-$. In the case of our model however, this channel is velocity suppressed, being the s-wave component of the cross section proportional to the electron mass. The leading order contribution is given therefore by the p-wave component, proportional to the DM velocity squared, $v_{\tilde{\chi}}^2$, which in the case of the GC is $v_{\tilde{\chi}} \sim 10^{-3}$. Nevertheless, as showed in~\cite{Bringmann:2012vr, Garny:2013, Giacchino:2013bta, Okada:2014zja, Garny:2015wea, Kopp:2014}, this helicity suppression can be lifted by the emission of a photon in the final state and, in particular, the virtual internal bremsstrahlung (VIB) yields a characteristic spectral feature that can resemble a monochromatic line and give information on the mass of DM. Another relevant channel is also the "smoking gun" process $\tilde{\chi} \tilde{\chi} \to \gamma \gamma$.
However, if we have a look at~\cite{Garny:2015wea} we can conclude that for small values of the mass splitting, $r \lesssim 2$, and in the DM mass range $10\; \textrm{GeV}<m_{\tilde{\chi}}< 10^4$ GeV the $\tilde{\chi} \tilde{\chi} \rightarrow e^- e^+ \gamma$ process dominates, being the only annihilation channel which matters.

Using the expressions for the annihilation cross section of the processes $\tilde{\chi} \tilde{\chi} \rightarrow e^- e^+ \gamma$ and $\tilde{\chi} \tilde{\chi} \rightarrow \gamma \gamma$ from the Appendix A of~\cite{Garny:2015wea}, in Fig.~\ref{fig:indirect} we show the current limits imposed by Fermi-LAT~\cite{fermi:2015lka} and HESS~\cite{hess:2018qtu} in the parameter space for the model points that provide the measured relic abundance. From the plot we can see that a portion of the parameter space at low DM mass is excluded by the Fermi-LAT limits, even if these constraints are still evaded in the scenario of small mass splitting, i.e. $\Delta M \lesssim 1$ GeV, and for $\Delta M \sim 10$ GeV. Above $m_{\tilde{\chi}} \sim 100$ GeV, however, the entirety of the points are still not probed by the current indirect detection experiments. The sudden change of color for points with $m_{\tilde{\chi}} \lesssim 50$ GeV starting at low mass with $\langle \sigma v \rangle \approx 3\cdot 10^{-30} \, \textrm{cm}^3/s$ is an artifact of the 2D projection of the space of solutions. In that region, there are always two  space points which return the same $\langle \sigma v \rangle$, one with small $\Delta M$ and one with big $\Delta M$. They have different values for the coupling but same value for $m_{\tilde{\chi}}$. Since they overlap, we decide to show the points which solve the relic density constraint with a higher mass splitting until only one solution is left.


It is important to mention that, in Fig~\ref{fig:indirect}, we limit ourselves to show parameter space points with $\Delta M > 10^{-1} \;\rm GeV$. This is motivated by the fact that, as it can be seen in Fig.~\ref{fig:relic}, we find solutions with $\Delta M < 10^{-1}  \;\rm GeV$ that satisfy the relic density requirements only above $m_{\tilde{\chi}}> 200$ GeV. However, the annihilation cross section $\langle \sigma v \rangle$ is basically insensitive to the value of the mass splitting and therefore they would just hide with the other solutions on the graph.



\subsection{Collider searches}
\label{subsec:collider}
Since the mediator of our model is colourless, it is difficult to produce at the LHC~\cite{Aad:2019qnd} and most of the constraints on the model come therefore from LEP data~\cite{lep:2005gc, Freitas:2014jla, lepdata}, see for example discussions in~\cite{Bringmann:2012vr, Kopp:2014, Garny:2015wea}. In particular, we can recast and reinterpret collider searches for right-handed slepton production, since our mediator particle is characterised by the same physical properties. 

In the scenario of small $m_\varphi$ (within the reach of the collider), mediators can be pair produced through EW interactions. If the mass splitting is moderate, i.e. the mediator is not long-lived, they subsequently decay in DM particles and electrons, yielding a characteristic same flavour opposite sign pair of leptons and missing transverse energy. However, if the mass splitting is particularly small (as in the cases under scrutiny in this work), the leptons would be extremely soft and therefore difficult to detect. Because of this, the LHC excludes only scenarios in which the mediator mass is $m_\varphi \lesssim 100$ GeV and $\Delta M \sim \textrm{few}$ GeV~\cite{Aad:2019qnd, Han:2014aea}. On the other hand, above right-handed slepton masses of $100$ GeV, the constraints are basically non existent in the compressed region. LEP constraints are more stringent and exclude pair production of sleptons with masses below $m_\varphi \sim 90$ GeV which only be evaded in the very degenerate scenario $r \leq 1.03$. At low mass, the strongest constraint comes from the $Z$ width measurement, which excludes sleptons with masses below $45$ GeV. Given that the coupling to the EW bosons is dictated only by gauge theory, if the mass of a slepton is below $M_Z/2$, a new decay mode would open up for the $Z$. Since none of these decay modes have been observed and the invisible width of the $Z$ is consistent with the expected neutrino contribution, that part of the mass spectrum is severely constrained independently of the coupling to the dark sector and the mass splitting. This therefore translates on a lower value for the mediator mass $m_\varphi$ that we will consider.

On the other hand, possible constraints from mono-photon events at LEP have been analysed in~\cite{Kopp:2014}. They find that these constraints are less stringent than the ones from Fermi LAT and HESS, which we have already shown in the previous section. We therefore do not take these into account.

\subsection{Direct detection}
\label{subsec:direct}
\begin{figure}[t!]
  \centering
  \includegraphics[width=.2\columnwidth]{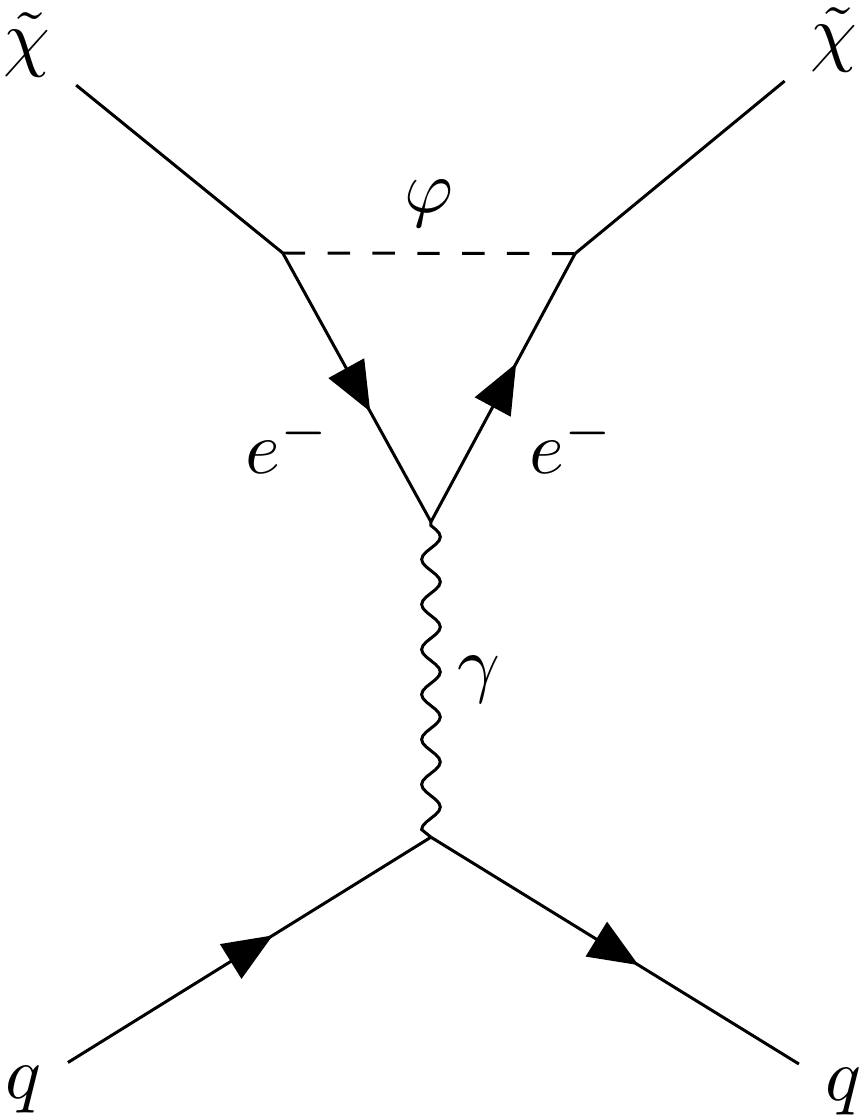} \qquad\qquad
  \includegraphics[width=.2\columnwidth]{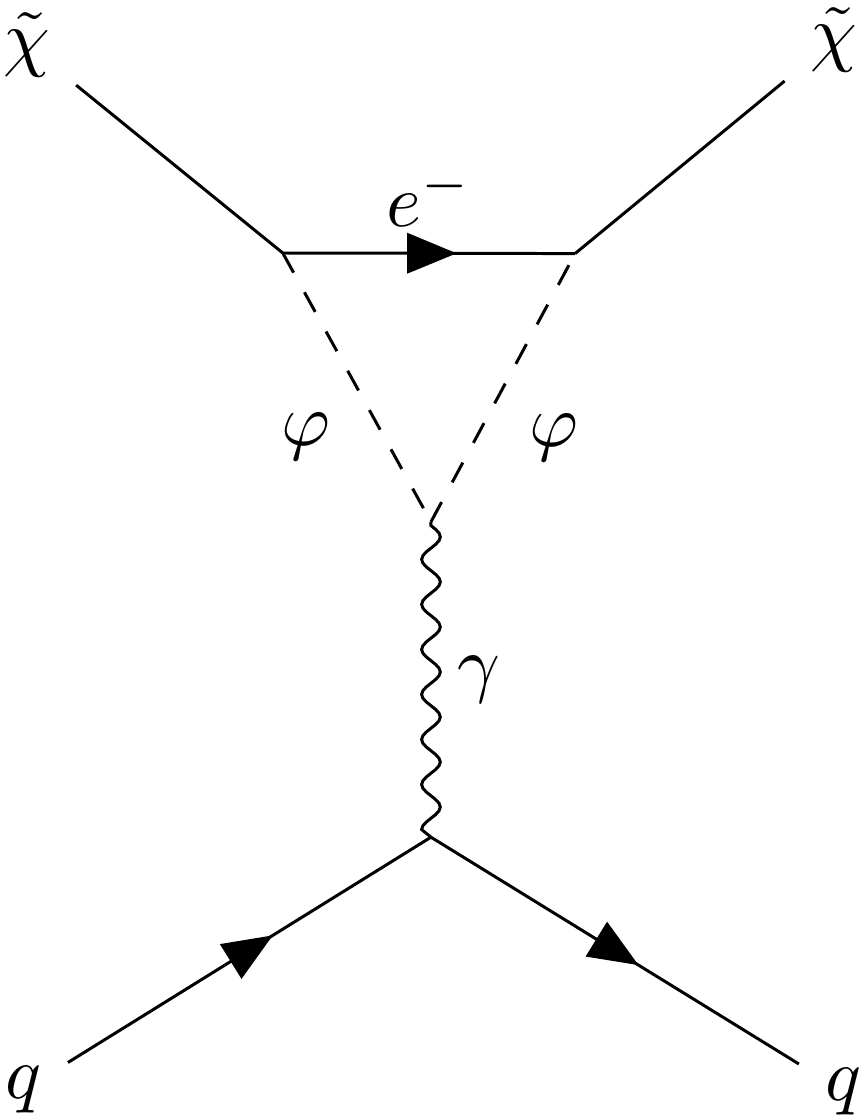}\
  \caption{
    \label{fig:diagrams_anapole}
     The one loop diagrams generating the effective DM–photon coupling for Majorana DM. Although not shown, two more diagrams with crossed two legs for $\tilde{\chi}$ are present.}
\end{figure}

\begin{figure}[!t]
  \centering
  \includegraphics[width=.6\columnwidth]{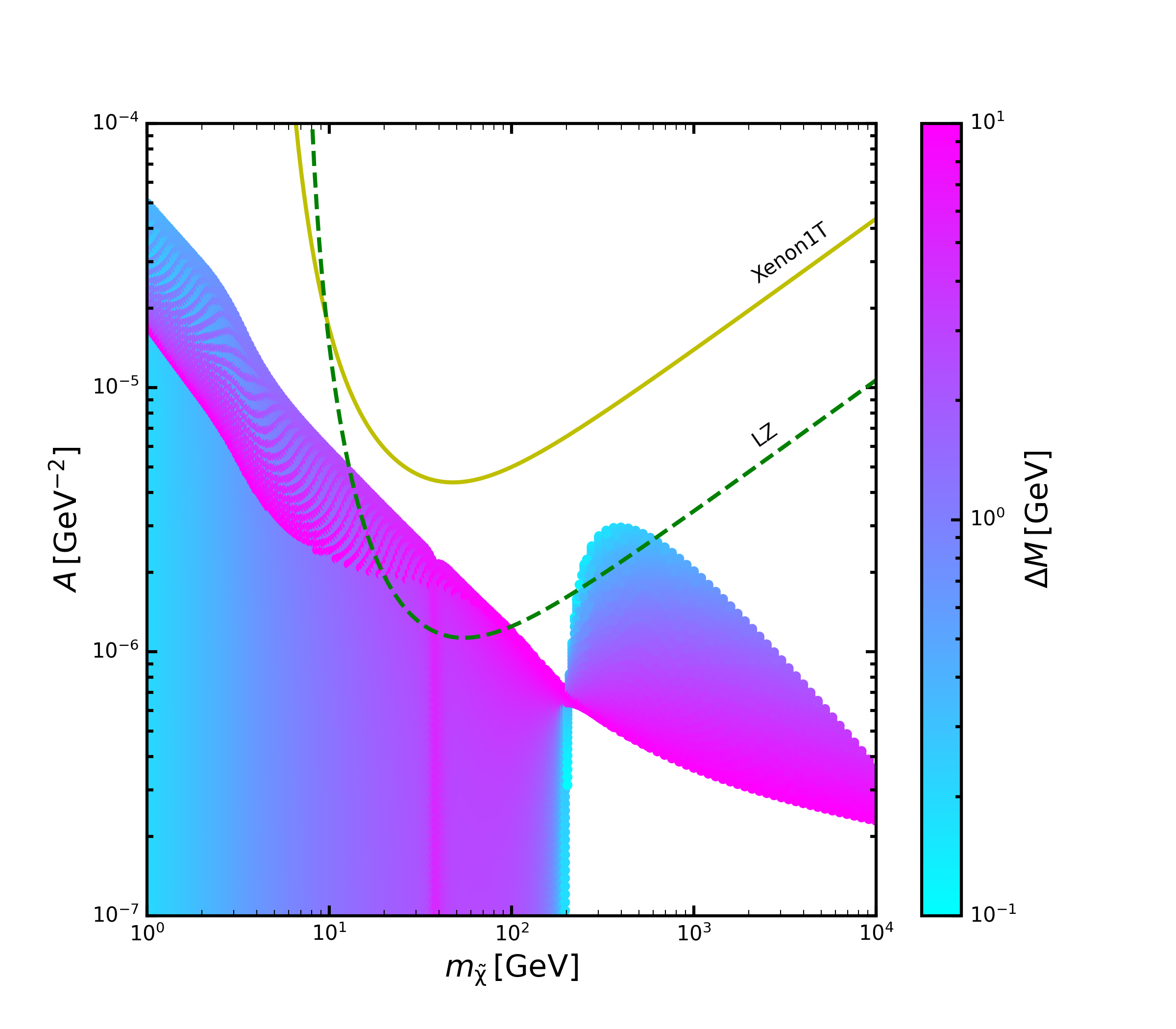}
  \caption{
    \label{fig:anapole}
    Anapole moment as a function of the DM mass for different values of the mass splitting which provide the correct relic density. We do not show cases with $r^2<1.001$, see the text for explanation. Solid yellow and dashed green lines provide the constraints from XENON1T~\cite{xenon1t:2018dbl} and the predicted reach of LZ~\cite{lz:2017qzi}, respectively. Data from~\cite{Cerdeno:2019vpd}.}
\end{figure}

Regarding direct detection searches, our Majorana DM particle can interact via one loop anapole moment with quarks as can be seen in Fig.~\ref{fig:diagrams_anapole}.

Following~\cite{Kopp:2014, Baker:2018uox}, the lagrangian for the effective vertex coupling to the photon through loops is given by
\begin{equation}
\mathcal{L}_{eff}= \mathcal{A} \bar{\psi}_{\tilde{\chi}} \gamma^\mu \gamma^5 \psi_{\tilde{\chi}} \partial^\nu F_{\mu \nu},
\end{equation}
with 
\begin{equation}
 \mathcal{A}= -\frac{e a_R^2}{32 \pi^2 m_{\tilde{\chi}}^2}\left[ \frac{-10+ 12\; \textrm{log} \left(\frac{\sqrt{|q^2|}}{m_{\tilde{\chi}}} \right)-(3+9 r^2)\; \textrm{log}(r^2-1)-(3-9r^2)\; \textrm{log}\; r^2}{9(r^2-1)}\right]
 \label{anapole}
\end{equation}
the anapole form factor with $r=m_{\varphi}/m_{\tilde{\chi}}$ for the limit in which the momentum transfer $q^2$ is higher than the mass of the particles in the loop (electrons in the case under study of this work), \textit{i.e.,} $|q^2| \gg m_e^2$. The transferred momentum of the interaction can be written as $\sqrt{|q^2|}=\sqrt{2 E_r m_T}$, where $E_r \sim \frac{1}{2}m_{\tilde{\chi}} v_{\tilde{\chi}}^2$ is the recoil energy and $m_T$ the mass of the target of the experiment. It is important to point out that there is a value of $r^2$ which depends on $m_{\tilde{\chi}}$ below which this perturbation theory approach is not valid anymore and the next term in the expansion should be added. To have an idea about the orders of magnitude we are talking about, for a DM mass of $50 \; \textrm{GeV}$ ($500 \; \textrm{GeV}$), this happens for values below  $r^2 \lesssim 1.001$ ($r^2 \lesssim 1.0001$). 

In Fig.~\ref{fig:anapole} we show the upper bounds given by XENON1T (solid)~\cite{xenon1t:2018dbl} and the projected ones from LZ (dashed)~\cite{lz:2017qzi} on the values of the anapole moment $\mathcal{A}$ for the parameter space points that provide the measured relic abundance. We can observe that XENON1T data cannot constrain our space of parameters while the expected sensitivity of LZ could exclude some values for $10\; \textrm{GeV}< m_{\tilde{\chi}} <10^3$ GeV. Since the calculation of the next term of the expansion for Eq. \eqref{anapole} is beyond the scope of this work, we only show the existent constraints for points of the space of parameters which are above $r^2 \sim 1.001$. 
Notice that also in this case we observe a sudden artificial step in solutions for $ m_{\tilde{\chi}} \lesssim 50$ GeV, which is the result of the 2D projection of the solutions, as explained for the indirect detection subsection.

Apart from direct detection, the anapole can be constrained in the context of collider searches. In particular, projected bounds from the High-Luminosity LHC can be seen in~\cite{Alves:2017uls}. We do not report here these limits, but it suffices to say that they are stronger than XENON1T but weaker than the projected LZ lines.

\section{Photon flux from cosmic ray scattering}
\label{sec:flux_results}
In this section we discuss the predictions for the flux of circularly polarised photons coming from the interaction of our DM candidate with CR electrons. In particular, the leading contribution comes from the $2 \to 3$ scattering $\tilde{\chi} e^- \rightarrow \tilde{\chi} e^- \gamma$.

It is worth mentioning that the possibility that a circular polarised signal is produced by these interactions was already pointed out in~\cite{Boehm:2017nrl}. In an attempt to illustrate this, they calculated the cross section for the electron neutralino scattering with a radiated positive and negative polarised photon in the final state for the best fit scenario in the MSSM and found that one cross section can be significantly greater than the other for some specific energy values. Based on this, they concluded that a net circular polarisation signal due to these interactions could be measured. Nevertheless, in order to determine if those circular polarised photons are detectable, the quantity that must be computed is the energy spectrum flux of circularly polarised photons from a known source. In this first computation of the spectrum, we focus on the GC because of its large DM density and its intense spectrum of electrons.

We define the flux of circularly polarised photons at a distance $r_\odot$ from the source, \textit{i.e.,} the GC, as

\begin{eqnarray}
\frac{d\Phi_{e\tilde{\chi}, pol}}{dE_\gamma}=\bar{J}\frac{1}{m_{\tilde{\chi}}}\int d\Omega_\gamma \int dE_e \frac{d\phi}{dE_e} \left|\frac{d^2\sigma_+}{d\Omega_\gamma dE_\gamma}(E_e,\theta_\gamma,E_\gamma )-\frac{d^2\sigma_-}{d\Omega_\gamma dE_\gamma}(E_e,\theta_\gamma,E_\gamma ) \right|,
\label{polflux1}
\end{eqnarray}
where $\Omega_\gamma$ is the solid angle between the emitted photon and the incoming CR electron (with $\theta_\gamma$ the polar coordinate), $E_e$ and $E_\gamma$ are the incoming electron and the outgoing photon energies, and $\bar{J}$ is the averaged J-factor that only depends on the spatial distributions of the astrophysical source. The $+$ and $-$ signs in the differential cross section indicate the positive and negative circular polarisations. The CR electron energy spectrum is described by $\frac{d\phi}{dE_e}$ and it can have a relevant impact on the overall flux and on the degree of net polarisation as well. We will discuss this in more details in the next section. Note that the total differential cross section can be expressed as
\begin{equation}
\frac{d^2\sigma}{d\Omega_\gamma dE_\gamma}(E_e,\theta_\gamma,E_\gamma )=\frac{d^2\sigma_+}{d\Omega_\gamma dE_\gamma}(E_e,\theta_\gamma,E_\gamma )+\frac{d^2\sigma_-}{d\Omega_\gamma dE_\gamma}(E_e,\theta_\gamma,E_\gamma ),    
\end{equation}
and the total flux 
\begin{equation}
\frac{d\Phi_{e\tilde{\chi}}}{dE_\gamma}=\bar{J}\frac{1}{m_{\tilde{\chi}}}\int d\Omega_\gamma \int dE_e \frac{d\phi}{dE_e} \frac{d^2\sigma}{d\Omega_\gamma dE_\gamma}(E_e,\theta_\gamma,E_\gamma ).
\label{totflux}    
\end{equation}
 Therefore, Eq. \eqref{polflux1} can be rewriten as
\begin{eqnarray}
\frac{d\Phi_{e\tilde{\chi}, pol}}{dE_\gamma}=\left|\frac{d\Phi_{e\tilde{\chi}, +}}{dE_\gamma}-\frac{d\Phi_{e\tilde{\chi}, -}}{dE_\gamma}\right|,
\label{polflux2}
\end{eqnarray}
where
\begin{eqnarray}
\frac{d\Phi_{e\tilde{\chi}, \pm}}{dE_\gamma}=\bar{J}\frac{1}{m_{\tilde{\chi}}}\int d\Omega_\gamma \int dE_e \frac{d\phi}{dE_e}  \frac{d^2\sigma_\pm}{d\Omega_\gamma dE_\gamma}(E_e,\theta_\gamma,E_\gamma )
\label{polfluxpm}
\end{eqnarray}
are the fluxes of positive and negative circularly polarised photons.

The spatial dependence of the DM and electron distributions are taken into account in the factor
\begin{eqnarray}
\bar{J}=\frac{1}{\Delta \Omega}\int_{\Delta \Omega} d\Omega J(\Omega)=\frac{2\pi}{\Delta \Omega}\int_{0}^{\theta_{\rm obs}} d\theta \; \textrm{sin}\theta J(\theta),
\label{jbar}
\end{eqnarray}
with
\begin{eqnarray}
J(\theta)=\int_0^{2r_\odot} ds \rho(r(s,\theta))f(r(s, \theta))
\label{jfactor}
\end{eqnarray}
the J-factor, which integrates over the line of sight the product of the DM density distribution, $\rho(r)$, and the function
\begin{equation}
f(r)=\frac{e^{-\frac{r}{r_0}}}{e^{-\frac{r_\odot}{r_0}}},
\label{fr}
\end{equation}
 which takes into account the fact that the CR flux is larger in the GC vicinity~\cite{Profumo:2011jt,Strong:2004td}. Here,  $r_\odot= 8.5$ kpc is the distance from the GC to the Earth and $r_0=4 \; \rm kpc$. Regarding the integration variables, $s$ is the radial distance from the Earth to the observed point of the target and $r$ the one from the center of the target (GC) to a given point inside it. In other words, $r^2=s^2+r_\odot^2-2r_\odot s \; \textrm{cos}\theta$, with $\theta$ the polar angle of observation. This quantity is averaged over the solid angle, $\Delta \Omega=2\pi (1-\textrm{cos}\; \theta_{\rm obs})$, where $\theta_{\rm obs}$ is the angular resolution of the experiment. See the Appendix B of~\cite{Gammaldi:2018} for details on the calculation of this quantity. 
We choose to focus on the direction towards the GC, where both DM and CR densities are higher. Given the typical angular resolution of Fermi-LAT~\cite{Fermi:2009} and e-ASTROGAM~\cite{DeAngelis:2017gra} we take $\theta_{\rm obs}= 1^{\circ}$ around the GC, which corresponds to $\Delta \Omega \sim 10^{-3}$. 
Regarding the DM density we consider a cuspy distribution, such as the conventional Einasto and NFW profiles. The parametrisation for these profiles are~\cite{einasto:1989,Navarro:1995iw}  
\begin{equation}
\rho_{\rm Ein}(r)= \rho_{s, \rm Ein} \; \textrm{exp} \left\{ -\frac{2}{\alpha}\left[ \left(\frac{r}{r_{s, \rm Ein}} \right)^\alpha-1 \right] \right\}  
\end{equation}
and 
\begin{equation}
\rho_{\rm NFW}(r)= \rho_{s, \rm NFW} \frac{r_{s, \rm NFW}}{r}\left(1+\frac{r}{r_{s, \rm NFW}}\right)^2,   
\end{equation}
respectively, where $\rho_{s, \rm Ein}=0.079 \; \rm GeV cm^{-3}$,  $r_{s, \rm Ein}= 20.0$ kpc and $\alpha=0.17$ are the parameters for the Einasto profile and $\rho_{s, \rm NFW}=0.307 \; \rm GeV cm^{-3}$ and $r_{s, \rm NFW}= 21.0$ kpc~\cite{Abramowski:2011hc,Navarro:2008kc} the ones related to the NFW distribution. The density profiles are normalized to the local DM density $\rho_0 = 0.39 \; \rm GeV \; cm^{-3}$~\cite{Catena:2009mf} at a distance $r_\odot= 8.5$ kpc from the GC.

\subsection{The Narrow Width Approximation}
\label{subsec:nwa}
\begin{figure}[t!]
  \centering
  \includegraphics[width=.48\columnwidth]{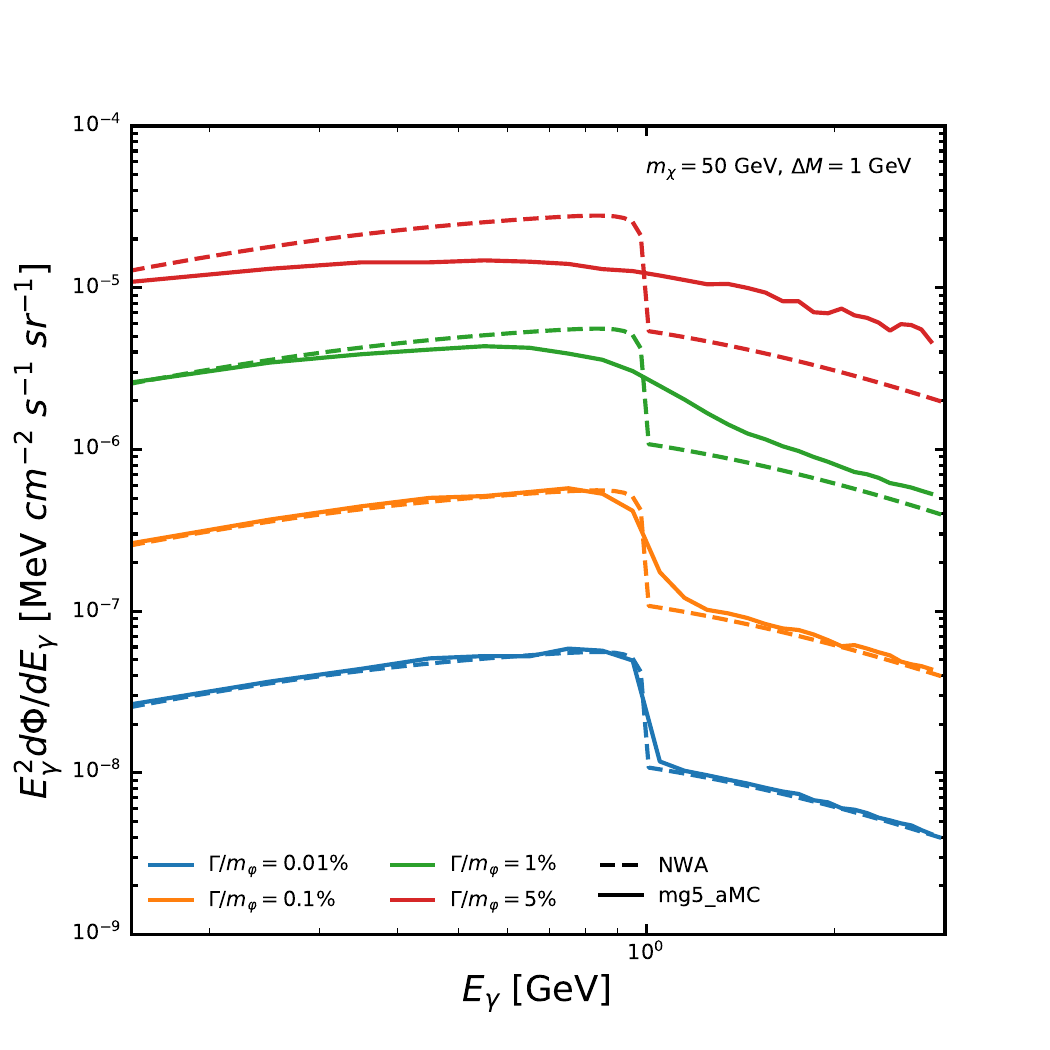}
  \includegraphics[width=.48\columnwidth]{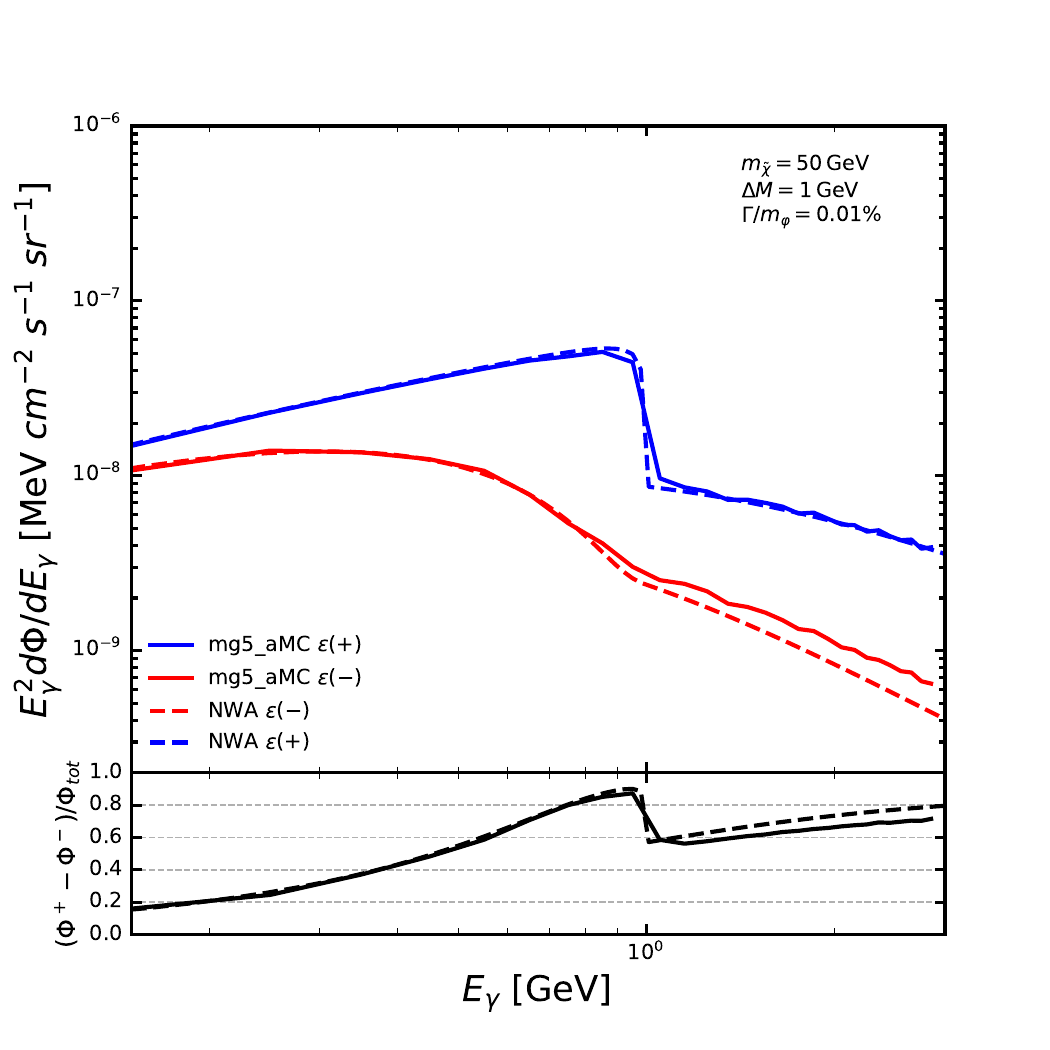}
  \caption{
    \label{fig:nwa}
    In the left plot: The total photon flux spectrum from CR electrons scattering off DM for $m_{\tilde{\chi}}=50$ GeV and $\Delta M=1$ GeV. Solid and dashed lines denote the results for the numerical calculation performed using a dedicated version of \mgFull and the one using the NWA respectively. We consider four values for $\Gamma/m_\varphi$ in different colours, i.e.. $a_R$ is chosen to  ensure these width to mass ratios. 
    In the right plot: The spectrum flux of positive (blue) and negative (red) circularly polarised photons using a dedicated version of \mgFull (solid line) and with the NWA calculation (dashed line) for $m_{\tilde{\chi}}=50$ GeV, $\Delta M=1$ GeV and $\Gamma/m_\varphi=10^{-4}$. In the lower panel we show the difference between the flux of positive and negative polarised photons over the total one. For both plots we have considered the Einasto DM density profile and $\frac{d\Phi}{dE_e}=k_e \left( \frac{E_e}{\rm GeV}\right)^{-3}$.
    }
\end{figure}



The process of interest has a kinematic feature that is helpful to discuss before getting down to the calculation. As a matter of fact, for certain electron initial energies, the process is resonant. Being interested in exploiting this feature to study the compressed region of the mass spectrum, we neglect the $u$ and $t$-channel diagrams in $e^- \tilde{\chi} \to e^- \tilde{\chi} \, \gamma$, after having verified that they are irrelevant in the parameter space region under study. Only the three resonant diagrams in Fig.~\ref{fig:diagrams} are retained.
We perform the calculation in the DM rest frame and denote $p_{\tilde{\chi}}=(m_{\tilde{\chi}}, \vec{0})$, $p_e=(E_e, \vec{p}_e)$ as the incoming DM and electron four-momenta, while $p'_{\tilde{\chi}}=(E^\prime_{\tilde{\chi}}, \vec{p}^{\, \prime}_{\tilde{\chi}})$, $p^\prime_e=(E^\prime_e, \vec{p}^{\, \prime}_e)$ are the outgoing ones. As it has been previously explained in~\cite{Gorchtein:2010xa}, the differential cross section is characterised by two resonances. The first and most important is the one in which $s=(p_e+p_{\tilde{\chi}})^2=m_\varphi^2$. The incoming electron energy that triggers the resonant production is fixed uniquely by the masses of the mediator and DM. In the mass degenerate scenario $m_{\tilde{\chi}} \approx m_\varphi$, we can write
\begin{equation}
E_{R1}=\frac{m_\varphi^2-m_{\tilde{\chi}}^2}{2m_{\tilde{\chi}}} \label{ER1} = \frac{(m_\varphi-m_{\tilde{\chi}})(m_\varphi+m_{\tilde{\chi}})}{2 m_{\tilde{\chi}}} \approx \Delta M \, ,
\end{equation}
indicating a direct relation between the mass splitting and the resonant initial electron energy. Given that the CR electron energy spectrum becomes fairly small beyond $10$ GeV, the above equation explains why the choice of $\Delta M \lesssim 10$ GeV is particularly sensible~\cite{Vittino:2019yme}. We will discuss this in more details later.
A second resonance is manifest when $s'=(p_e'+p_{\tilde{\chi}}')=m_\varphi^2$, which is given when the initial electron energy is
 \begin{equation}
     E_{R2}=\frac{m_\varphi^2-m_{\tilde{\chi}}^2+2m_{\tilde{\chi}} E_\gamma}{2(m_{\tilde{\chi}}-E_\gamma(1-\cos{\theta_\gamma}))}.
     \label{ER2}
 \end{equation}
Contrary to the first resonance, the second one is given by a continuum of initial electron energies, parametrised in the above expression by $\cos{\theta_\gamma}$.
Note that in both cases the mass of the electron $m_e$ has been neglected.

Since we are interested in cases in which $\Delta M$ is small with respect to the mediator mass, the total width of the mediator is given by
\begin{equation}
    \Gamma_{\rm tot} \sim \Gamma_{\varphi \rightarrow e^- \tilde{\chi}} = \frac{1}{8\pi}a_R^2\frac{(m_\varphi^2-m_{\tilde{\chi}}^2)^2}{2 m_\varphi^3} \approx \frac{1}{4\pi}a_R^2\frac{(\Delta M)^2}{m_\varphi}   \, .
\end{equation}
Therefore, the width to mass ratio is small,  
\begin{equation}
 \frac{\Gamma_{\rm tot}}{m_\varphi} \propto \left(\frac{\Delta M}{m_\varphi}\right)^2 \ll 1 \, .
\end{equation}
This suggests that we are in the ideal regime to employ the narrow width approximation (NWA)~\cite{Berdine:2007uv}, which allows us to factorise the production and the decay of the mediator. In order to verify the approximation and have under control the neglected off-shell effects, we validate the calculation by comparing the NWA flux with the one obtained with a dedicated version of \mgFull. This version allows for a user-defined energy profile for the initial state, making it possible to obtain the integration over the CR electron energy.
The numerical integration is made with the full matrix element and can be performed for any value of $\Gamma_{\rm tot}/m_\varphi$. Nevertheless, when $\Gamma_{\rm tot}/m_\varphi \ll 1$ the phase space become quite cumbersome and dedicated techniques are needed to help the Monte Carlo generator to converge. For this reason the NWA is preferred, allowing us to speed up the calculations while providing reliable results in the validity regime (see Fig.~\ref{fig:nwa} for the validation).

Once we focus on a specific region of the sky, $\bar{J}$ is fixed and the quantity we need to calculate is %
\begin{equation}
\frac{d\tilde{\Phi}_{e\tilde{\chi}, \pm}}{dE_\gamma}\equiv \frac{m_{\tilde{\chi}}}{\bar{J}} \frac{d\Phi_{e\tilde{\chi}, \pm}}{dE_\gamma}=  \int dE_e \frac{d\phi}{dE_e}  \frac{d\sigma_\pm}{ dE_\gamma}(E_e,E_\gamma). 
\label{fluxNWA}
\end{equation}
In the derivation of the NWA approximation we will refer to the first, second and third diagram considering their order in the lower part of Fig.~\ref{fig:diagrams}.
As in the NWA we factorise production and decay, when the propagator in the first diagram is on-shell, the propagators in the other two are not. This means that in this approximation there is no interference between the first diagram and the other two, allowing us to split the calculation in two pieces.
We rewrite Eq. \eqref{fluxNWA} as
\begin{equation}
\frac{d\tilde{\phi}_{e\tilde{\chi}, \pm}}{dE_\gamma}=\int \frac{d\phi}{dE_e}\frac{d\tilde{\sigma}_{1\pm}}{dE_\gamma} \textrm{d}E_e+\int \frac{d\phi}{dE_e} \frac{d\tilde{\sigma}_{2\pm}}{dE_\gamma} \textrm{d}E_e, 
\label{fluxnwa}
\end{equation}
where
\begin{equation}
\frac{d\tilde{\sigma}_{1\pm}}{dE_\gamma}=\sigma_{e^- \tilde{\chi} \rightarrow \varphi}(E_e) \frac{d\Gamma_{\varphi \rightarrow e^- \tilde{\chi} \gamma^\pm}}{dE_\gamma}(E_\gamma)  \frac{1}{\Gamma_{\rm tot}} 
\label{eq:nwafirst}
\end{equation}
is the differential cross section for the first diagram in the NWA and 
\begin{equation}
\frac{d\tilde{\sigma}_{2\pm}}{dE_\gamma}= \frac{d\sigma_{e^- \tilde{\chi} \to \varphi \gamma^{\pm}}}{dE_\gamma} (E_e, E_\gamma)   \frac{\Gamma_{\varphi \rightarrow e^- \tilde{\chi}}}{\Gamma_{\rm tot}}
\label{nwasecond}
\end{equation}
is the differential cross section for the other two, with $\Gamma_{\rm tot}=\Gamma_{\varphi \rightarrow e^- \tilde{\chi}}+\Gamma_{\varphi \rightarrow e^- \tilde{\chi} \gamma^\pm}$ the total decay width of the mediator. Note that in first approximation $\Gamma_{\rm tot} \sim \Gamma_{\varphi \rightarrow e^- \tilde{\chi}}$. In doing so, we describe the process in two parts: one characterised by a $2 \to 1$ production mode and a $1 \to 3$ decay, and the other by a $2 \to 2$ production mode and a 2 particles decay.
The $2 \to 1$ production mode is given by
\begin{equation}
\sigma_{e^- \tilde{\chi} \rightarrow \varphi}(E_e)=\frac{\delta(E_e-E_{R1}) \pi}{4\sqrt{(E_e m_{\tilde{\chi}})^2-m_{\tilde{\chi}}^2 m_e^2}}\frac{|\overline{\mathcal{M}}|_{e^- \tilde{\chi} \rightarrow \varphi}^2}{E_e+m_{\tilde{\chi}}} \, ,
\end{equation}
while the differential 3 particle decay width reads
\begin{equation}
\frac{d\Gamma_{\varphi \rightarrow e^- \tilde{\chi} \gamma^\pm}}{dE_\gamma}(E_\gamma) =\frac{1}{8(2\pi)^3 m_\varphi} \int_{E_{e, min}'}^{E_{e, max}'}   |\overline{\mathcal{M}}|_{\varphi \rightarrow e^- \tilde{\chi} \gamma_\pm}^2dE_e',  
\end{equation}
with $E_{e, min}'$ and $E_{e, max}'$ the minimum and maximum energies which satisfy the kinematic constraints for a given $E_\gamma$.
Note that, although the 3 body decay should be considered in flight, for small mass splitting values, this effect is negligible since
\begin{equation}
   \frac{E_\varphi}{m_\varphi}= \frac{E_{R1}+m_{\tilde{\chi}}}{m_\varphi}\sim 1 
\end{equation}
and we therefore compute the decay at rest.
The last missing term is the $2 \to 2$ differential cross section in Eq.~\eqref{nwasecond}
\begin{equation}
 \frac{d\sigma_{e^- \tilde{\chi} \rightarrow \varphi \gamma^{\pm}}}{dE_\gamma} (E_e, E_\gamma) =\frac{E_\gamma}{32\pi\sqrt{(E_e m_{\tilde{\chi}})^2-m_e^2m_{\tilde{\chi}}^2}}\int dx\; \frac{ |\overline{\mathcal{M}}|_{e^- \tilde{\chi} \rightarrow \varphi \gamma^{\pm}}^2 }{E_e+m_{\tilde{\chi}}-|\vec{p_e}|x}\frac{\delta(x-x^0)}{\left|\frac{\partial E_\gamma}{\partial x}\right|_{x^0}} \, ,
\end{equation}
where $x = \cos{\theta_\gamma}$ and
%
 \begin{equation}
x^0=\frac{2E_e(E_\gamma-m_{\tilde{\chi}})+2E_\gamma m_{\tilde{\chi}}+m_\varphi^2-m_{\tilde{\chi}}^2-m_e^2}{2E_\gamma |\vec{p}_e|}.    
 \end{equation}
It is worth pointing out that thanks to the resonant enhancement, the cross section is not proportional to $a_R^4$ as naively expected, but to $a_R^2$. Looking for example at Eq.~\eqref{eq:nwafirst} or Eq.~\eqref{nwasecond}, it is indeed self-evident that this is the case by simply checking the coupling dependence of each term.

In the left plot of Fig.~\ref{fig:nwa} we show the comparison of the results for the total photon flux spectrum obtained from the numerical integration (solid lines) and the one we get with the NWA approximation (dashed lines). Following~\cite{Profumo:2011jt}, an electron energy spectrum given by 
\begin{equation}
    \frac{d\Phi}{dE_e}=k_e \left( \frac{E_e}{\rm GeV}\right)^{-3} \, ,
\end{equation}
with $k_e= 10^{-2} \; \rm GeV^{-1} \; cm^{-2} \; s^{-1}\; sr^{-1}$ has been considered. We assume an Einasto DM density profile and choose a benchmark mass point $m_{\tilde{\chi}}=50$ GeV and $\Delta M=1$ GeV, while we vary the width to mass ratio to test the validity of the approximation. We observe that the NWA is a good estimation when $\Gamma_{\rm tot}/m_\varphi \lesssim 0.1\%$ but the neglected off-shell effects start to play a relevant role above $1\%$. 

On other hand, on the right side of Fig.~\ref{fig:nwa} we show for the same benchmark mass point the decomposition of the flux in negative and positive circular polarisations. We compare once again the results with the numerical calculation, setting $\Gamma_{\rm tot}/m_\varphi = 0.1\%$. In the inset we can see the difference between the positive and negative polarised photon fluxes over the total one, i.e.
\begin{equation}
    A_{pol} = \frac{\Phi^+-\Phi^-}{\Phi_{\rm tot}} \equiv \frac{\frac{d\Phi_{e\tilde{\chi}, pol}}{dE_\gamma}}{\frac{d\Phi_{e\tilde{\chi} }}{dE_\gamma}} \, .
\end{equation}
In terms of the asymmetry $A_{pol}$ the polarisation fraction of the flux can be understood as
\begin{equation}
    F_{pol} = \frac{\Phi^+}{\Phi_{tot}} = \frac{1 + A_{pol}}{2} \, .
\end{equation}
Note that when the mass splitting is low enough, $\Gamma/m_\varphi \lesssim 0.1\%$, there is a sharp drop-off in the photon energy spectrum beyond the kinematic threshold for which the first resonance is not allowed. This happens when the minimum electron energy,
\begin{equation}
    E_{\rm min}= \frac{E_\gamma}{1-\frac{E_\gamma}{m_{\tilde{\chi}}}\textrm{cos}\; \theta_\gamma},
\end{equation}
is higher than $E_{R1}$, i.e. 
\begin{equation}
    E_\gamma > \frac{m_{\tilde{\chi}}(m_\varphi^2-m_{\tilde{\chi}}^2)}{2m_{\tilde{\chi}}^2+(m_\varphi^2-m_{\tilde{\chi}}^2)(1-\textrm{cos}\; \theta_\gamma)} \, .
\end{equation}
This means that the drop-off will happen between $E_{\gamma, 1}=\frac{m_{\tilde{\chi}}(m_\varphi^2-m_{\tilde{\chi}}^2)}{2m_\varphi^2}$ and $E_{\gamma, 2}=\frac{(m_\varphi^2-m_{\tilde{\chi}}^2)}{2m_{\tilde{\chi}}}$. For $\Delta M \ll m_{\tilde{\chi}},\;m_\varphi $, this takes place for $E_\gamma \sim \Delta M$. Beyond this energy, only the second resonance contributes to the flux.
See~\cite{Gorchtein:2010xa} for more details on the kinematics.
Since the photon is attached to a right-handed electron in the first diagram, leading to the production of the first resonance, P is violated and an excess of positive polarised photons is generated.
Specifically, this excess  is maximal in the sharp edge of the flux, where the positive polarisation is subject to an enhanced energy behaviour.
 
\subsection{Results}
\label{subsec:results}
\begin{figure}[t!]
  \centering
  \includegraphics[width=.48\columnwidth]{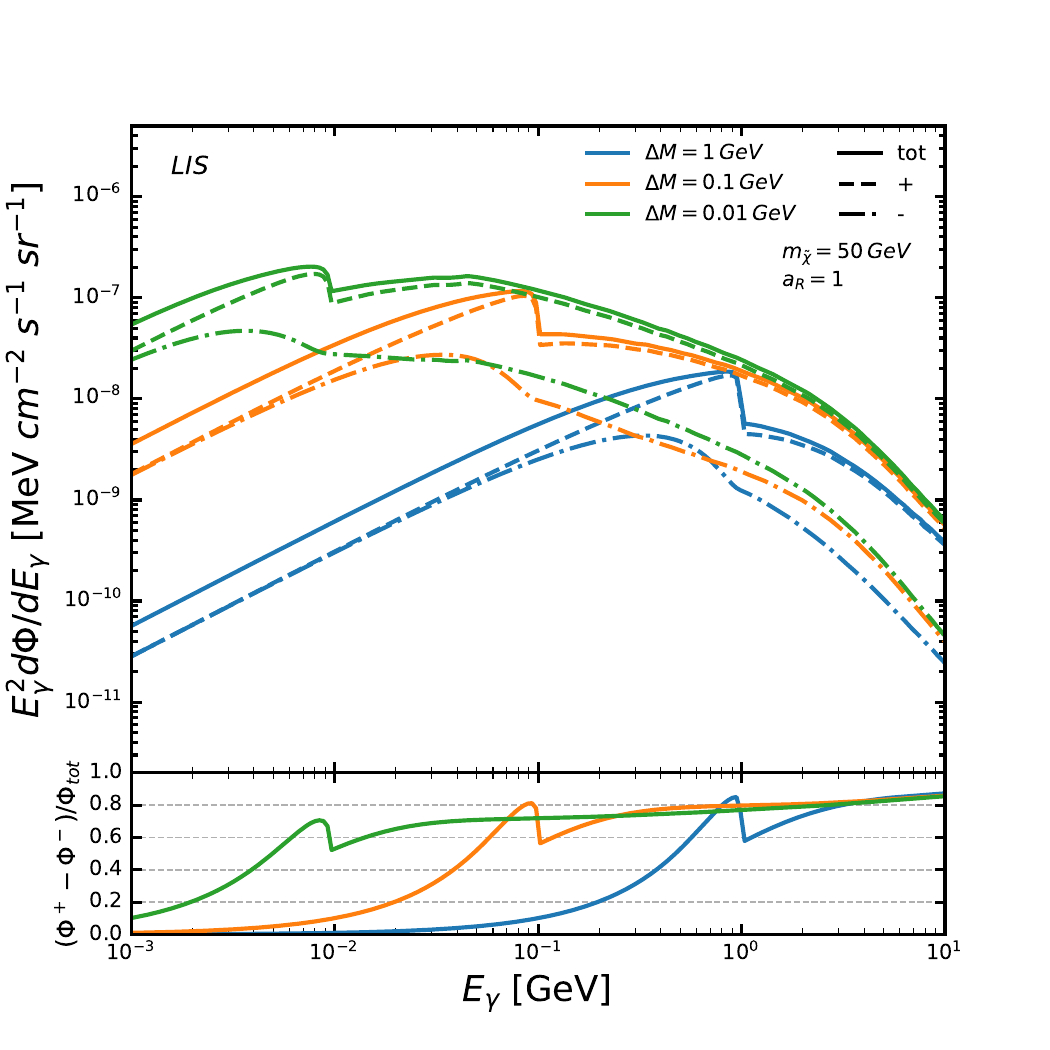}
  \includegraphics[width=.48\columnwidth]{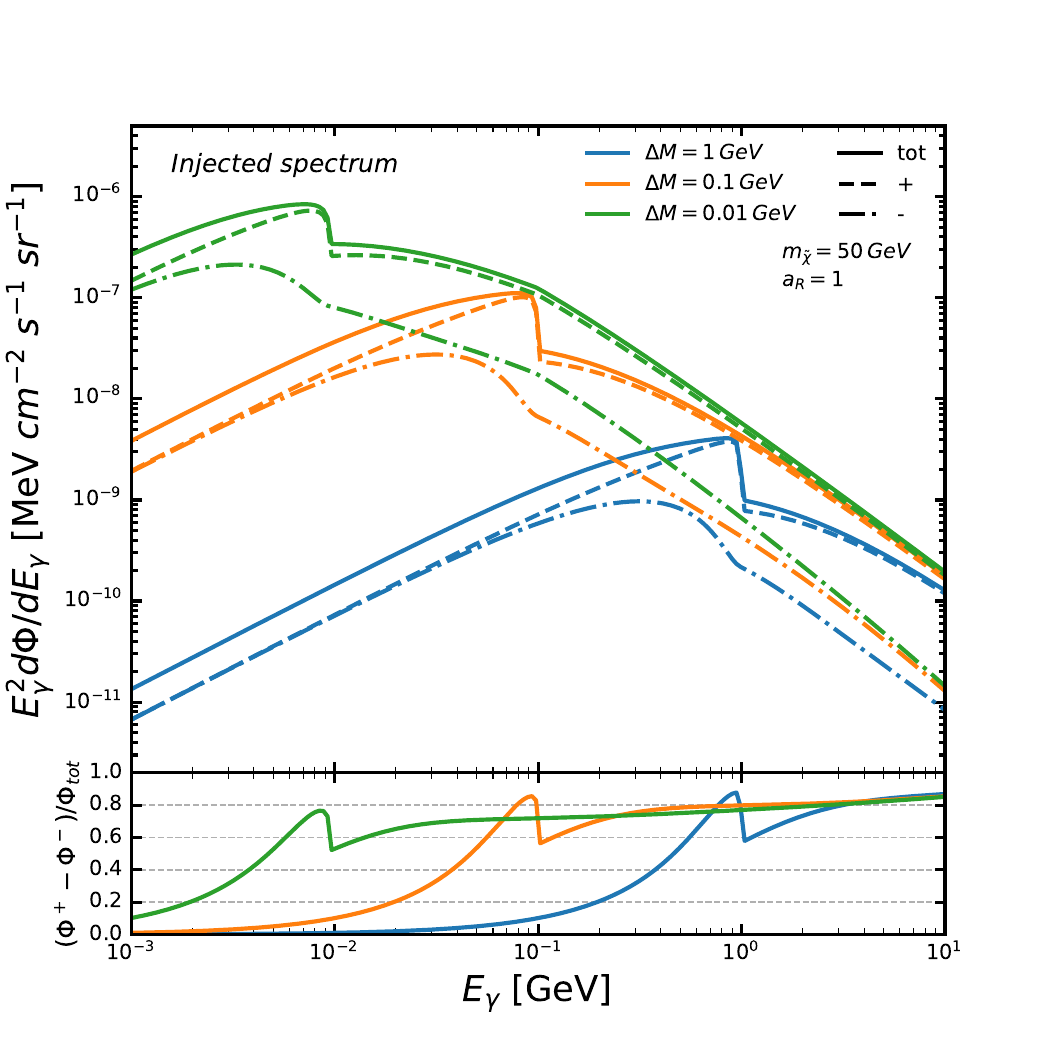}\\
  
  \includegraphics[width=.48\columnwidth]{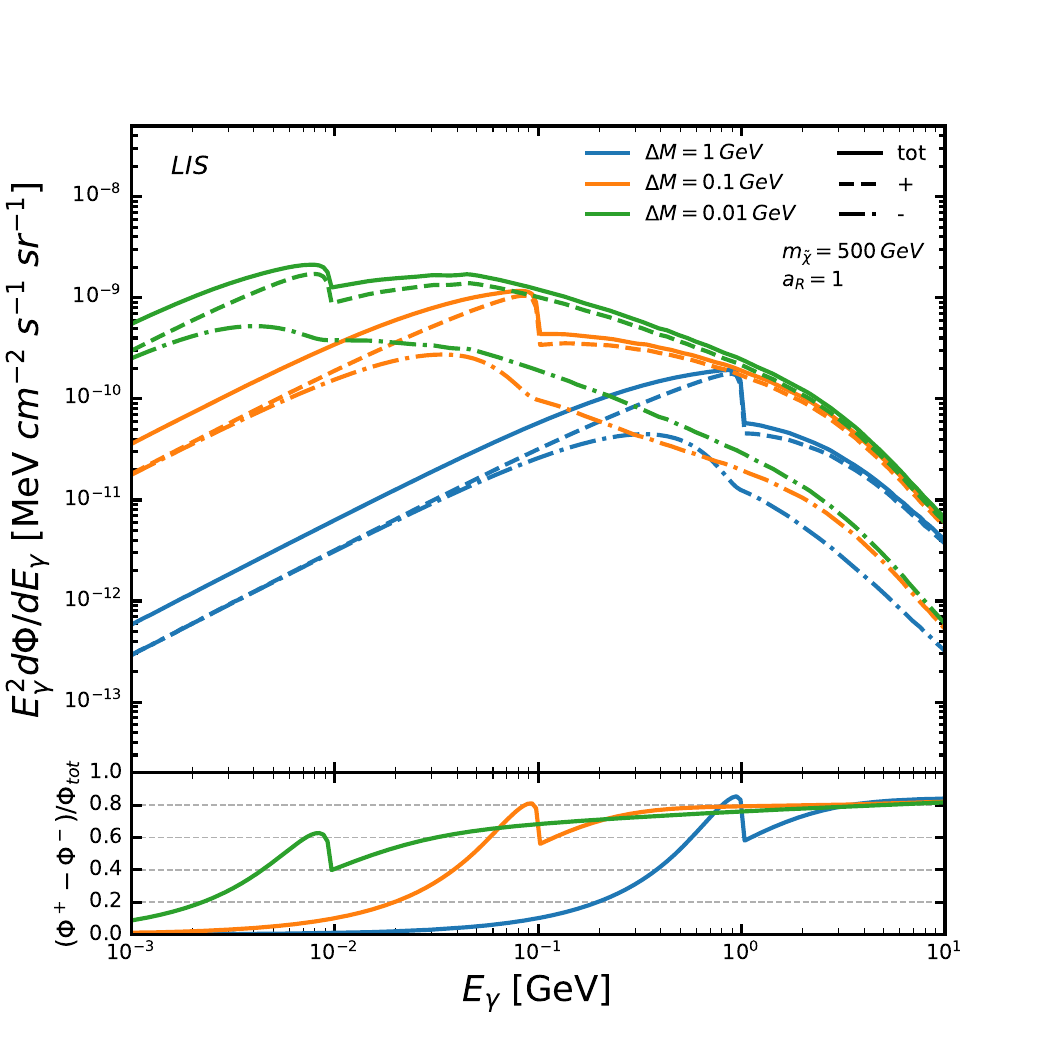}
  \includegraphics[width=.48\columnwidth]{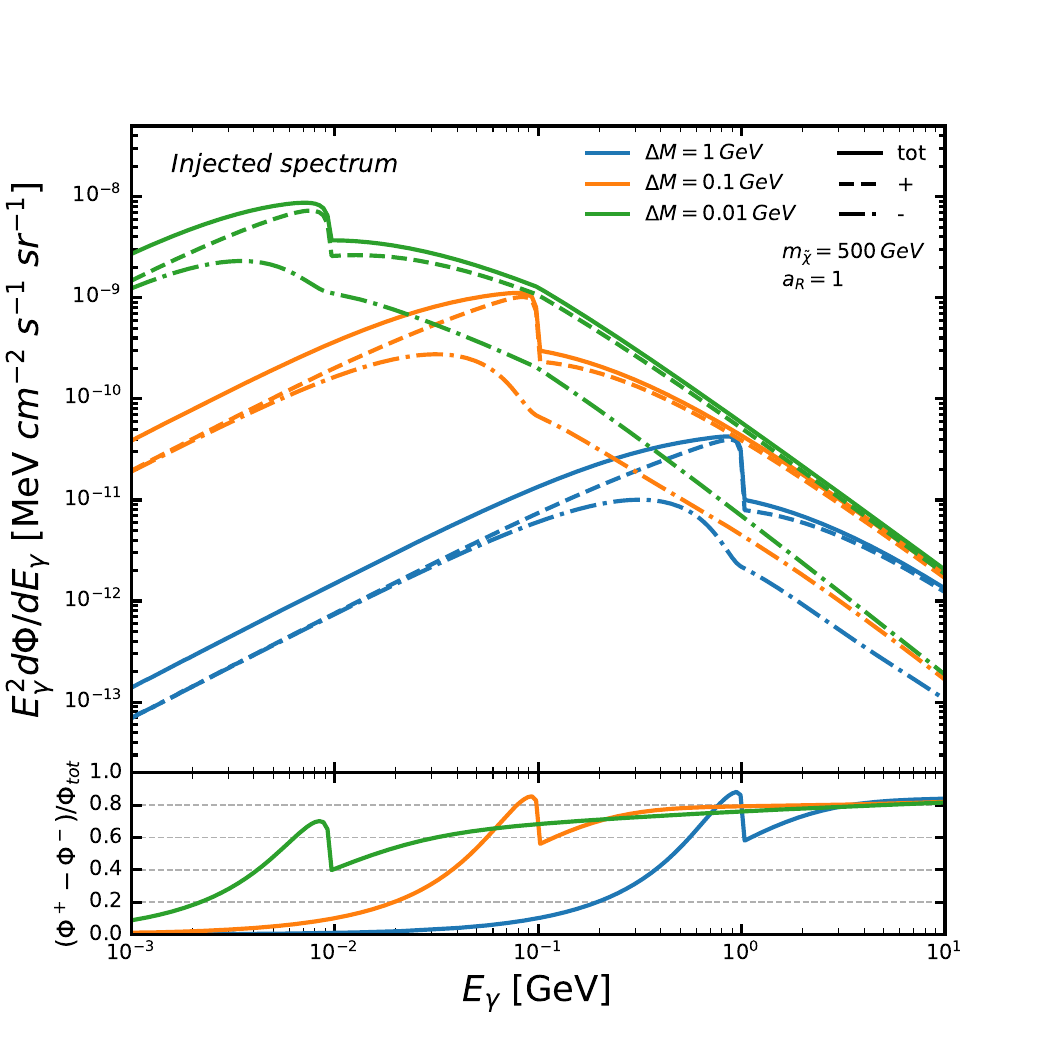}
  \caption{
    \label{fig:flux}
  The flux spectrum of negative and positive circularly polarised photons in dash-dotted and dashed lines, respectively, and the sum of them, in solid lines, from the scattering of CR electrons off DM. We show in green, orange and blue the results obtained for $\Delta M=1$ GeV $0.1$ GeV and $0.01$ GeV. In the lower part of each plot the difference between the flux of positive and negative polarised photons over the total one, i.e the asymmetry $A_{pol}$, can be seen.
  In the left panel: the local interstellar spectrum for electrons expected at Earth derived in~\cite{Vittino:2019yme} has been used. In the right panel: the injected spectrum of electrons given by~\cite{Vittino:2019yme} has been considered.
  We show the cases of $m_{\tilde{\chi}}=50$ GeV (upper plots) and $m_{\tilde{\chi}}=500$ GeV (bottom plots).
  We set $a_R=1$ and consider the Einasto DM density profile.    }
\end{figure}

\begin{figure}[t!]
  \centering
  \includegraphics[width=.48\columnwidth]{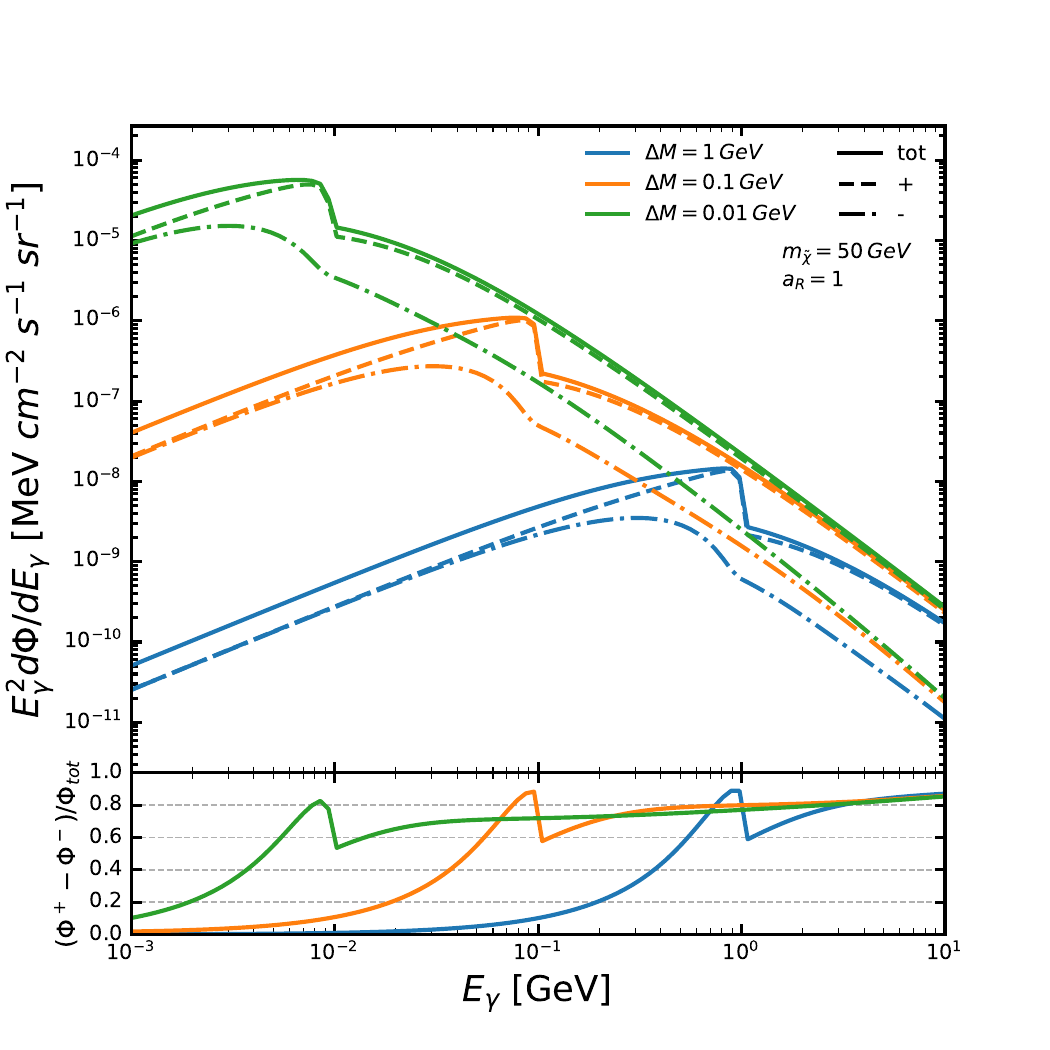}
  \includegraphics[width=.48\columnwidth]{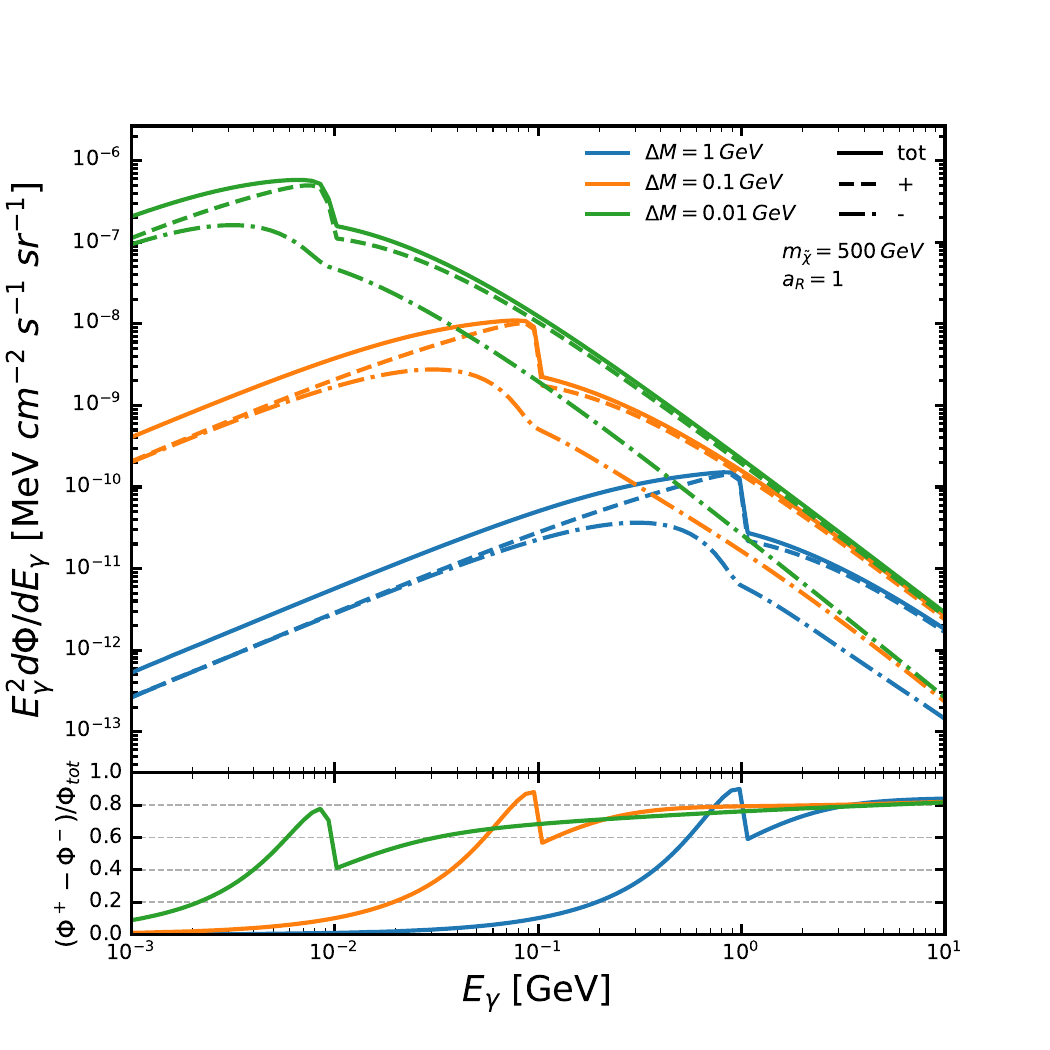}
  \caption{
    \label{fig:flux2}
  The flux spectrum of negative and positive circularly polarised photons, dash-dotted and dashed lines respectively, and the sum of them, solid lines, from the scattering of CR electrons off DM. We show in green, orange and blue the results obtained for $\Delta M=1$ GeV, $0.1$ GeV and $0.01$ GeV, respectively. In the lower part of each plot the difference between the flux of positive and negative polarised photons over the total one can be seen.
  We consider the Einasto DM density profile, $\frac{d\Phi}{dE_e}=k_e \left( \frac{E_e}{\rm GeV}\right)^{-3}$ and set $a_R=1$. }
\end{figure}

Once verified that the NWA is a good estimate for the cases of interest of this work, we use this approximation to study the flux of polarised photons coming from DM scattering with CR electrons in the GC. 

In the following we show the results assuming the Einasto DM density profile. However, if we wanted to consider the NFW DM density profile, we would only need to multiply our results for the flux by $\frac{\bar{J}_{\rm NFW}}{\bar{J}_{\rm Ein}} \sim 0.7$ (since $\bar{J}_{\rm NFW}=1.19 \; 10^{24}\; \rm GeV/ cm^2$ and $\bar{J}_{\rm Ein}=1.67 \; 10^{24}\; \rm GeV/ cm^2$). 

From this point on, we relax the requirement that the DM particle is a thermal relic which provides the total relic abundance. This is motivated by the fact that this calculation is also applicable for candidates which are produced by non-thermal mechanisms.

In Fig.~\ref{fig:flux} we show the flux spectrum of negative and positive circularly polarised photons and their sum (dash-dotted, dashed and solid lines respectively). We fix the DM mass to $m_{\tilde{\chi}}=50$ GeV (upper panel) and $m_{\tilde{\chi}}=500$ GeV (bottom panel), the coupling $a_R=1$ and take different values of $\Delta M$, namely $1$, $0.1$ and $0.01$ GeV. In the lower part of each plot the difference between the flux of positive and negative polarised photons over the total one, i.e. the circular polarisation asymmetry $\mathcal{A}_{pol}$, is reported.
As naively expected, we observe that the higher the DM mass and $\Delta M$ are, the lower the flux is. 

The energy spectrum for the CR electrons used in~\cite{Profumo:2011jt} is only a good approximation for the one measured at Earth for $E_e \sim 5-10^3$ GeV, see~\cite{Ackermann:2010ij}. As we are interested in lower energies, from MeV to a few GeV, we extract the energy spectrum data from Fig. 9 of~\cite{Vittino:2019yme} and we compute the integrals of Eq. \eqref{polfluxpm} using an interpolation of these data. The results of this are shown on the left panel of Fig.~\ref{fig:flux}. 
Just to have an idea of the behaviour of this energy spectrum, we have checked that it scales as
\begin{align}
\nonumber
    E_e^{-1.2} \quad &\textrm{for} \quad E_e < 0.05 \; \textrm{GeV} \, ,\\
    E_e^{-2} \quad &\textrm{for} \quad 0.05 \; \textrm{GeV} \lesssim E_e \lesssim 4 \; \textrm{GeV} \, , \label{eq:espectrum} \\
    \nonumber
    E_e^{-3} \quad &\textrm{for} \quad E_e > 4 \; \textrm{GeV} \, .
\end{align}
Note that for $E_e \sim 10^{-1}-10$ GeV, due to the large effect of solar modulation (plus effects of a geomagnetic cut-off), the local interstellar spectrum cannot be directly measured and it is derived by assuming an injected spectrum based on a particular model which fits the data measured at Earth by Voyager at low energies and by AMS at $E_e \gtrsim 10$ GeV~\cite{Vittino:2019yme}.
Although a full-fledged CR and gamma-ray analysis is needed to take into account how the energy spectrum of the electrons changes during their travel from the GC to Earth, this analysis is beyond the scope of this work. 

On the right panel of Fig.~\ref{fig:flux} we can see the results that we get if instead we consider the electron injected spectrum, \textit{i.e.}, the spectrum of electrons injected by supernova remnants into the interstellar medium. In order to be consistent with the local interstellar spectrum interpolated from data of~\cite{Vittino:2019yme}, we use the injected spectrum of the $1$ break model
\begin{equation}
\frac{d\Phi_{\rm inj}}{dE_e} =
    \begin{cases}
      k_{e, \textrm{inj}} \left(\frac{E_e}{\textrm{GeV}}\right)^{-2.13}  & \text{for $E_e \leq 0.109$ GeV}\\
     \frac{k_{e, \textrm{inj}}}{8.9 \; 10^{-3}} \left(\frac{E_e}{0.109 \; \textrm{GeV}}\right)^{-2.57}  & \text{for $E_e > 0.109$ GeV}
    \end{cases} 
    \label{eq:injspectrum}
\end{equation}
with $ k_{e, \textrm{inj}}=6.98 \; 10^{-3} \; \rm GeV^{-1} \; cm^{-2} \; s^{-1}\; sr^{-1}$.

Looking at Fig.~\ref{fig:flux} we can conclude that the asymmetry of polarised photons coming from these interactions can reach up to $90\%$. The highest value for the flux for a fixed value of $\Delta M$ is obtained at $E_\gamma=\Delta M$ but this is not necessarily true for the polarisation fraction (see the bottom part of the plots). For $\Delta M =0.01$ GeV, we can see that the highest values of $A_{pol}$ are obtained for $E_\gamma \gg \Delta M$, where the flux decreases significantly, whereas for higher $\Delta M$ the value of $A_{pol}$ at $E_\gamma=\Delta M$ is similar to the constant value reached for $E_\gamma \gg \Delta M$. This feature depends on the energy spectrum of the electrons, meaning that if the energy spectrum is the same for all the energy range, we do not observe these differences. As an example of this, in Fig.~\ref{fig:flux2} we plot the same cases that we consider in Fig.~\ref{fig:flux} but fixing the energy spectrum to $\frac{d\Phi}{dE_e}=k_e \left( \frac{E_e}{\rm GeV}\right)^{-3}$. We can see there that as the electron energy spectrum is the same for all the energy range, the circular polarisation fraction shape does not change for different values of the mass splitting, contrary to the case of Fig.~\ref{fig:flux}. Consequently, a measurement of the polarisation asymmetry could help us gaining insight on the electron CR spectrum.

Additionally, both from Fig.~\ref{fig:flux} and Fig.~\ref{fig:flux2}, we can notice that the smaller $\Delta M$ is, the lower the circular polarisation asymmetry becomes, contrary to what happens with the intensity of the flux. This is more notable in Fig.~\ref{fig:flux}, and is ultimately related to the fact that the electron energy spectrum for energies around $10^{-2}$ GeV is less steep than at energies $\sim 10^{-1}$ GeV and above. This can be verified by looking at the power law behaviours in Eq.~\ref{eq:espectrum} and Eq.~\ref{eq:injspectrum} for the injected case.
In order to understand why this is the case, we computed the asymmetry separately for the two components of the flux, i.e. the first and second resonance contributions. We found that the polarisation asymmetry when we only consider the diagram which contributes to the first resonance (first diagram of Fig.~\ref{fig:diagrams}) is independent of the parameter space point and always yields a maximal polarisation fraction of $100\%$ on the peak situated at $E_\gamma \sim \Delta M$. On the other hand, the second resonance is characterised in general by a much lower polarisation asymmetry, as a consequence of the fact that the photon from the third diagram in the second row of Fig.~\ref{fig:diagrams} is emitted from the scalar mediator. Contrary to the electron, the mediator, does not remember that it has participated in a P violating interaction because it has no spin, and therefore that diagram does not produce polarised photons, polluting the signal and lowering the total polarisation fraction in the first resonance peak. It can be verified that the second resonance is relatively more important at smaller $\Delta M$ and this motivates the fact that the asymmetry decreases for lower values of $\Delta M$. In addition, because of the lower slope of the LIS and the injected electron spectrum for energies around $0.01$ GeV, it is more likely that second resonance photons are produced at the peak of the flux for $\Delta M = 0.01$ GeV, lowering the polarisation asymmetry considerably with respects to the other two scenarios considered. If instead we use a spectrum with a shape which is scale invariant such as $ \sim E^{-3}$, see Fig.~\ref{fig:flux2}, the relative contribution of the second resonance is still bigger for lower values of $\Delta M$ but not at the same level than in Fig.~\ref{fig:flux}. On the other hand, comparing the results of the left and right plots of Fig.~\ref{fig:flux}, we can see that for any value of $\Delta M$, the asymmetry is higher when we consider the injected spectrum over the LIS. This is because, due to its higher slope in the whole energy range considered, the injected spectrum favours energies which populate the first resonance over the ones which populate the second. This effects is even more enhanced in the case of the $E^{-3}$ spectrum profile.

From the differences found in our results when changing the electron energy profile from the local interstellar to the injected one, we can conclude that the electron energy spectrum has a relevant impact both on the circular polarisation asymmetry and on the flux.
The realistic spectrum however will be something in between the two limit cases considered here. 
Therefore, we can expect that for a realistic full-fledged simulation both the flux and the circular asymmetry can have significant variations, retaining nonetheless certain basic features that are direct consequence of the kinematics.

It should be noted that, if we understand our DM candidate as a thermal relic, the benchmark examples shown in the plots are not providing the total relic density and the flux (not the asymmetry) need to be rescaled by the appropriate factor $a_R^2$, see Fig.~\ref{fig:relic}.
In the case of $m_{\tilde{\chi}}=50$ GeV, no solution was found for the relic density at low mass splitting and the expected thermal production would therefore be underabundant.

In addition to the discussed signatures of circularly polarised photons at $E_\gamma \sim \Delta M$, we would expect another peak of unpolarised photons around $E_\gamma \sim m_{\tilde{\chi}}$ due to self-annihilation. In order to naively assess the scale of these processes, we can estimate the annihilation rate as
\begin{equation}
    q_{\tilde{\chi}\tilde{\chi}} \sim \left<\sigma v\right> \frac{\rho_0}{m_{\tilde{\chi}}} \sim \frac{a_R^4 e^2}{(4 \pi)^2 m^2_{\tilde{\chi}}} v_{\tilde{\chi}} \frac{\rho_0}{m_{\tilde{\chi}}} \, .
\end{equation}
 On the other hand, the CR scattering rate can be written as
\begin{equation}
    q_{e\tilde{\chi}} \sim \sigma_{e\tilde{\chi}} E_e \frac{d\phi}{dE_e} \sim \frac{a_R^2 e^2}{4 \pi m_{\tilde{\chi}}^2} E_e \frac{d\phi}{dE_e} \, ,
\end{equation}
where the DM-electron scattering cross section benefits from a lower order coupling dependence thanks to the resonant enhancement.
If we take the ratio of the two, we find
\begin{equation}
\frac{q_{e\tilde{\chi}}}{q_{\tilde{\chi}\tilde{\chi}}}  \sim  10^{-6} a_R^{-2} \left(\frac{E_e\frac{d\phi}{dE_e}}{\rm cm^{-2}\; s^{-1}} \right)\left(\frac{m_{\tilde{\chi}}}{\rm GeV}\right) \, ,
\end{equation}
which implies that at small couplings and high masses the CR scattering process can become relevant. It is however important to remember that the two spectra are complementary since they populate the photon spectrum at very different energies, one yielding information on the mass splitting $\Delta M$ and the other on the mass of DM.

Finally, it is interesting to note that these results are also valid for the case of Dirac DM. In this scenario, while the DM particle interacts through the t-channel, the scattering with the antiparticle is potentially s-channel enhanced. Because of this, in order to get the corresponding flux one should rescale the above predictions by the ratio between the antiDM and the total DM abundances. If DM particles are asymmetric~\cite{Petraki:2013wwa, Zurek:2013wia}, i.e. there is an asymmetry between the number of particles and antiparticles, annihilation is not efficient and DM-CR electron scattering would be the only process producing a signal in the gamma ray band. This is for instance the case study discussed in~\cite{Profumo:2011jt}. Nevertheless, unless antiparticles dominate over particles, the process would not be resonant and the flux is expected to be negligible compared with the one obtained in this work.

\section{Prospects of detection}
\label{sec:detection}

The detection of a polarised signal of the kind previously described could potentially help us gain knowledge on both DM and CR spacial and energy distributions. Although no existing experiment has been designed to measure the circular polarisation of photons in the gamma-ray band, in the scenario of a dedicated future experiment, the possible astrophysical sources of background have to be taken into account.

In order to analyse the detectability of DM from the measurement of circularly polarised photons, we must have in mind the standard processes which can generate circular polarisation since they could act as a background for our signal. The main sources of circularly polarised photons are
\begin{description}[font=$\bullet$\scshape\bfseries]
\item Faraday conversion~\cite{Bavarsad:2009hm, Haghighat:2019rht}: converts linear polarisation to circular polarisation when passing through a high magnetic field.
\item Birefringence~\cite{Montero-Camacho:2018vgs}: linear polarisation which passes through a medium of aligned grains whose alignment twists along the line of sight can be transform into circular polarisation.
\item Synchrotron emission and curvature radiation~\cite{Gangadhara_2010, deBurca:2015kea}: intrinsic origin. Both linear and circular polarisation can be generated.
\end{description}

Here, we consider DM-CR electron interactions happening in the GC, a high DM density area from which we do not expect any peak of circular polarisation in the gamma-ray band apart from the one predicted in this work. Although radio synchrotron radiation is observed from the GC due to the surrounding magnetic fields, gamma-rays are believed to be produced by other processes such as inverse-Compton scattering, bremsstrahlung of electrons scattering off ambient photons or gas and inelastic collisions of high energy CRs (mostly protons) with the ambient medium (producing neutral pions which in turn decay)~\cite{vanEldik:2015qla}. On the other hand, important linear polarisation signals are not expected either, suggesting that the possibility of Faraday conversion~\cite{Bavarsad:2009hm, Haghighat:2019rht} can be neglected. The source of gamma-rays from the GC is still not clear. Different candidates are the supermassive black hole Sgr A$*$, the supernova remnant Sgr A East, the pulsar wind nebulae G359.95-0.04 or thousand millisecond pulsars which could be present in the central star cluster surrounding Sgr A*~\cite{Eckner:2017oul, Cholis:2018izy}. See~\cite{vanEldik:2015qla} for a review. However, none of these sources would produce gamma-rays via synchrotron or curvature radiation, meaning that these photons are not expected to be circularly polarised. 

It is worth mentioning that circular polarisation can be generated due to hadronic collisions in cosmic accelerators or in the atmosphere as long as there is an excess of protons over antiprotons in the initial state. However, since the circular polarisation fraction due to these processes is at the level of $5 \; 10^{-4}$ \cite{Boehm:2019yit}, it is negligible compared with the expected signal from a maximally P violating DM interaction.

In Section~\ref{sec:flux_results} we have shown that the flux of photons coming from the DM-CR electron interaction under study can provide a peak in the spectrum of photons with a circular polarisation asymmetry reaching up to $90\%$. Given the fact that we do not expect any other bump in the spectrum of photons with a significant level of circular polarisation from astrophysical sources in the energy band under consideration, the measurement of this highly polarised photon flux could open a new window for DM exploration. 

Moreover, there is possibility that the polarisation could change during the journey from its emission point to Earth. This problem was addressed in~\cite{Boehm:2019lvx}, finding that while this is a possible effect, it depends on the characteristic energy of the photon spectrum. In particular, the phenomenon appears to be negligible for photon energies bigger than $10$ MeV. We therefore do not discuss further the matter and assume that the net circular polarisation is completely preserved in the propagation through the galaxy medium.

Having all of these in mind, we analyse the experimental prospects for detecting a net circular polarisation from the GC. 
First of all, we estimate the sensitivity needed for a typical counting experiment like Fermi-LAT~\cite{Fermi:2009} or e-ASTROGRAM~\cite{DeAngelis:2017gra} to measure the total flux of photons coming from the interactions under consideration. In order to do this, we need to take into account the astrophysical background of photons coming from the region of interest (ROI) that the experiment is looking at. In this context, we consider the diffuse background energy flux in the energy range $1\; \textrm{MeV} \leq E_\gamma \leq 1\; \textrm{GeV}$ for a ROI of $10^{\circ} \times 10^{\circ}$ ($\Delta \Omega=0.03$ sr) predicted by~\cite{Gaggero:2017dpb} (Fig. 2 black solid line), which is $E_\gamma^2\frac{d\phi_{\rm back}}{dE_\gamma} \sim 10^{-2} \; \rm MeV \; cm^{-2} \; s^{-1} \; sr^{-1} $.

Defining $N_{\rm signal}$ as the number of photons of the signal that we expect to measure and $N_{\textrm{back}}$ as the ones coming from the astrophysical background, we can find an estimated sensitivity in order to have a 3 $\sigma$ discovery. In particular, for a given energy resolution $\epsilon$, an effective area $A_{\textrm{eff}}$ and a time exposure $\Delta t$ of the detector, for a ROI characterized by a solid angle $\Delta \Omega$, we can estimate   
\begin{equation}
\frac{N_{\rm signal}}{\sqrt{N_{\textrm{back}}}} \sim  \frac{\frac{d\Phi_{e\chi}}{dE_\gamma}}{\sqrt{\frac{d\Phi_{\textrm{back}}}{dE_\gamma}}}\sqrt{2 \epsilon E_{\gamma, \textrm{peak}}\Delta \Omega A_{\textrm{eff}} \Delta t } = 3\, ,  
\label{sign}
\end{equation}
where $E_{\gamma, \textrm{peak}} \sim \Delta M$ is the energy corresponding to the peak of our signal.
Since e-ASTROGAM is expected to be able to measure the photon flux in the whole energy range under study and it will be capable of measuring the linear polarisation of the signal, we consider its instrumental details
\begin{align*}
    \epsilon=0.013 \, , \quad A_{\textrm{eff}}= 50-560 \; \rm cm^2 \quad &\mathrm{for} \quad E_\gamma = 0.3-10 \,
  \mathrm{MeV}\\
  \epsilon=0.3 \quad A_{\textrm{eff}}=  215-1810
\; \rm cm^2 \quad &\mathrm{for} \quad E_\gamma= 10-3000 \,
  \mathrm{MeV} \, .
\end{align*}
In this way, we find that, for a time exposure of $\Delta t \sim 10^8$ s, the order of magnitude of the needed signal is 
\begin{equation}
    E_\gamma^2\frac{d\Phi_{e\chi}}{dE_\gamma} \sim 10^{-5} \; \rm MeV\;  cm^{-2}\; s^{-1}\; sr^{-1} \, .
\end{equation}

Note that, although this estimation has been done for an experiment like e-ASTROGAM, it can be observed that a similar sensitivity is obtained if we consider the instrument details of Fermi-LAT, which collects data in the energy range $20$ MeV to $300$ GeV.
 
We remind the reader that the results shown in Figs.~\ref{fig:nwa},~\ref{fig:flux} and~\ref{fig:flux2} were derived for an angular region of $\Delta \Omega \sim 10^{-3}$ sr ($\theta_{\rm obs}=1^{\circ}$). For this ROI, the average J-factor takes the value $\bar{J}_{\theta_{\rm obs}=1^{\circ}}=1.67 \; 10^{24} \; \rm GeV/cm^2$ ($\bar{J}_{\theta_{\rm obs}=1^{\circ}}=1.19 \; 10^{24} \; \rm GeV/cm^2$ for the NFW profile). 
Nevertheless, for a correct comparison with the background, we need to calculate $\bar{J}$ for the same ROI where the background flux has being modeled, i.e. $10^{\circ} \times 10^{\circ}$. In galactic coordinates, $|l| \leq 5^{\circ}$ and $|b| \leq 5^{\circ}$, see for example~\cite{Bell:2020rkw}, we obtain $\bar{J}_{10^{\circ} \times 10^{\circ}}=9.02 \; 10^{23} \; \rm GeV/cm^2$ ($\bar{J}_{10^{\circ} \times 10^{\circ}}=6.35 \; 10^{23} \; \rm GeV/cm^2$ for NFW). For a fair comparison, we must therefore multiply our results by the factor $\frac{\bar{J}_{10^{\circ} \times 10^{\circ}}}{\bar{J}_{\theta_{\rm obs}=1^{\circ}}} \sim 0.54$ in order to quantify how far our fluxes are from the sensitivity of the experiments. Even if we consider the highest signal found for the injected spectrum, in the 1 break model considered, we would still need an exposure time of $\Delta t \sim 10^{10}$~s for a $3 \sigma$ evidence. However, it is important to note that for an angular resolution of $\theta_{obs}=1^{\circ}$ the background would be much smaller than the for $10^{\circ} \times 10^{\circ}$, increasing the sensitivity to measure our signal.

It is noteworthy that, although results for DM masses lower than the ones considered in Section~\ref{subsec:results}, i.e. $m_\chi \sim 5$ GeV, would provide fluxes which reach the sensitivity of e-ASTROGAM for $\Delta M \sim 0.01$ GeV in $\Delta t \sim 10^{8}$ s, we do not show these results here since these masses are excluded by the constraint of the $Z$ boson decay, $m_\varphi \geq 45$ GeV, as already discussed in Section~\ref{subsec:collider}. Even though this constraint has not been taken into account in some past works like \cite{Profumo:2011jt, Kopp:2014, Okada:2014zja}, it cannot be easily evaded since in the simplified model considered the mediator must be $U(1)_Y$ charged in order to preserve gauge invariance. There is no straightforward way of decoupling the $Z$ boson, unless exotic UV assumptions are made concerning a full gauge invariant theory. If that was the case, we could argue that the flux of photons coming from the interaction of a $5$~GeV DM candidate with CR electrons could be measured in the first year of operation of e-ASTROGAM.
 
If instead of measuring the signal itself we focus on the circular polarisation fraction, we have to be aware of the fact that no efficient methods using non-Compton scattering techniques for gamma-ray circular polarimetry have been developed to date, limiting the energy range of detectable photons to $0.3-30$ MeV. In~\cite{Elagin:2017cgu} it is argued that, due to the fact that circular polarisation techniques measure the secondary asymmetries to determine the polarisation of a primary gamma-ray flux, the sensitivity will decrease by a factor $A \sqrt{\epsilon}_p$, where $A \sim 10\%$ is the typical asymmetry in the secondary particle spectra expected from a $100\%$ polarised gamma-ray flux and $\epsilon_p$ the efficiency of the detector for useful events.
 In the best case scenario in which the exposure time is enough to measure the signal, since expected efficiencies for useful events are $\epsilon_p \simeq 10^{-3}-10^{-4}$ (for the method involving Compton scattering of magnetized iron) and $\epsilon_p \simeq 10^{-6}$ (for the method using bremsstrahlung asymmetry of Compton scattered electrons of unpolarised matter), we find necessary for new techniques to be developed if we want to be able to measure the circular polarisation of photons coming from the interactions considered in this work. 
To the best of our knowledge, no obvious experimental technique is available to date to exploit significant polarisation fractions with the objective of increasing sensitivity to the signal. If such methods were available, a new window in indirect detection could be opened, expanding our potential to discover and characterise DM.

Finally, it is important to remark that although here we have focused on the signals coming from the GC, a different source of CR particles could be considered in order to maximize the total photon flux and improve the detection prospects. For example, the interaction of DM with electrons and protons close to cosmic accelerators such as active galactic nuclei, pulsars and supernova remnants can also provide circularly polarised photons. As these charged particles are accelerated to very high energies around these objects, in addition to the fact that their density as well as the DM one is known to be enhanced with respect to the one considered here, we can expect a higher flux of photons compared to the results of this work. In that case, if the background is moderate enough, the observation prospects might be more encouraging. This further study is however left for a future work. Furthermore, another direction to explore is to consider different P violating models. In case of a magnified photon flux, the measurement of the high circular polarisation asymmetry predicted in this work could become feasible and help us understanding the nature of DM interactions.

\section{Conclusions}
\label{sec:conclusions}

In this work, we have discussed the importance of the photon polarisation as a mean to characterise and potentially detect DM interactions.
In particular, if the DM interaction with SM particles is P violating and an asymmetry is present for the number density of particles over anti-particles of one of the initial states, a distinctive circular polarised signal can be generated from the scattering of CRs and DM particles.

In order to provide quantitative results, we have considered a simplified t-channel model in which parity is maximally violated and a fermionic DM couples to right-handed electrons by means of a charged scalar mediator. In particular, we have focused on the low mass splitting region of the parameter space in order to exploit the resonant enhancement of the model when incoming electrons collide with DM particles in high DM density regions. After discussing the current constraints on the model posed by the most recent experimental results, we have computed for the first time the differential flux of circular polarised photons coming from the GC, highlighting that the expected polarisation asymmetry of the signal can reach up to $90\%$ at the peak observed in the energy spectrum when the photon energy takes the value of the mass splitting. Both the photon flux and the circular polarisation asymmetry are highly dependent on the mass splitting and the electron energy spectrum, suggesting that different sources, such as cosmic accelerators (active galactic nuclei, pulsars, supernova remnants, etc.), or different models could produce higher circular polarised fluxes and potentially lead to the detection of DM. We leave the study of these cases for a future work.

Finally, we have discussed the detection prospects in the near future by experiments like Fermi-LAT and e-ASTROGAM. Although a distinctive peak will be present in the photon flux spectrum, the signal obtained does not seem to be detectable in the immediate future, unless novel techniques are devised to reduce backgrounds and to improve the angular resolution and the sensitivity to the polarisation fraction. Another potential way to have higher photon fluxes would be to consider lighter DM and mediator candidates, such as $m_\chi \sim 5$ GeV, since fluxes are proportional to the inverse power of $m_\chi^2$. Unfortunately, an important constraint on the model, which is often neglected, is the one coming from the $Z$ boson decay when the mass of the mediator is less than $45$ GeV. This forces us to consider heavier mediator and DM masses which lead to suppressed fluxes. Unless unconventional UV assumptions are made to avoid the coupling to the $Z$ or a different P violating model is considered, the only reasonable option to obtain a higher intensity flux is from different sources where CR electrons are more abundant or present a different energy profile.



\section*{Acknowledgments}
We thank L. Ubaldi, S. Robles,  J. Heisig, D. G. Cerde\~no, J. A. Aguilar-Saavedra, A. Cheek and S. Profumo for helpful discussions and comments. The work of C.D. and M.C. was funded by the F.R.S.-FNRS through the MISU convention F.6001.19. The work of L.M. has received funding from the European Union’s Horizon 2020 research and innovation programme as part of the Marie Sklodowska-Curie Innovative Training Network MCnetITN3 (grant agreement no. 722104). Computational resources have been provided by the supercomputing facilities of the Universit\'e Catholique de Louvain (CISM/UCL) and the Consortium des \'Equipements de Calcul Intensif en F\'ed\'eration Wallonie Bruxelles (C\'ECI) funded by the Fond de la Recherche Scientifique de Belgique (F.R.S.-FNRS) under convention 2.5020.11 and by the Walloon Region.


\biboptions{sort&compress}
\bibliographystyle{Science.bst}
\bibliography{refs}


\end{document}